\documentclass{article}
\usepackage{amssymb}

\usepackage{graphicx}
\usepackage{amsmath}

\newtheorem{theorem}{Theorem}
\newtheorem{acknowledgement}[theorem]{Acknowledgment}

\newtheorem{assumption}[theorem]{Assumption}

\newtheorem{corollary}[theorem]{Corollary}

\newtheorem{definition}[theorem]{Definition}

\newtheorem{lemma}[theorem]{Lemma}

\newtheorem{proposition}[theorem]{Proposition}
\newtheorem{remark}[theorem]{Remark}

\newenvironment{proof}[1][Proof]{\textbf{#1.} }{\ \rule{0.5em}{0.5em}}

\newcommand {\X}{{\cal X}}
\newcommand {\F}{{\cal F}}
\renewcommand {\L}{{\cal L}}

\newcommand {\N}{{\cal N}}
\newcommand {\W}{{\cal W}}
\newcommand {\Q}{{\cal Q}}
\renewcommand {\O}{{\cal O}}

\newcommand{\susy}{\mathfrak{susy}}
\renewcommand{\Im}{\mbox{{\rm Im}}}

\newcommand{\Sym}{\mbox{{\rm Sym}}}
\newcommand {\Ker}{\mbox{{\rm Ker}}}

\newcommand{\dbar}{\bar \partial}
\newcommand{\nequiv}{\equiv\!\!\!\!\!\!/}

\newcommand{\ctym}{{\cal TYM}}

\begin{document}
\title{Deformation of maximally supersymmetric Yang-Mills theory in dimensions 10. \\ An algebraic approach.} 
\author{M.Movshev \thanks{The work was partially supported by grants DE-FG02-90ER40542  and PHY99-0794}\\ IAS \\Princeton\\NJ USA}
\date{\today}
\maketitle
\begin{abstract}
We make a preliminary algebraic study of supersymmetric  deformations of $N=1$ Yang-Mills theory in dimension ten with an arbitrary gauge group. This is done in a context of Lie algebra deformation theory. The tangent space to the space of deformation is computed.
\end{abstract}

\section{Introduction}
In this paper we shall study the following problem. Suppose
\begin{equation}\label{E:YM}
{\cal L}(\nabla,\xi) =<F_{ij},F_{ij}>+\Gamma_{\alpha\beta}^i<\nabla_i\xi^{\alpha},\xi^{\beta}>
\end{equation}
is a Lagrangian of supersymmetric Yang-Mills theory in dimension ten. Let $\theta_{\alpha}$ be supersymmetry transformations. 

Some problem in physics can be formulated as a deformation problem: to find a certain deformation
\begin{equation}\label{E:dsfdjh1}
{\cal L}_{\alpha'}(\nabla,\xi) ={\cal L}(\nabla,\xi)+\sum_{k\geq 1}{\alpha}^{'k}{\cal L}^k(\nabla,\xi)
\end{equation}
where ${\alpha'}$ is a parameter of deformation. The deformed supersymmetry transformations 
\begin{equation}
\theta_{\alpha }({\alpha'}) =\theta_{\alpha }+\sum_{k\geq 1}\theta^k_{\alpha }{\alpha'}^k
\end{equation}
should leave the action corresponding to the density \ref{E:dsfdjh1} invariant. The Lie algebra generated by $\theta_{\alpha }({\alpha'})$, restricted to space of critical points of ${\cal L}_{\alpha'}$ is isomorphic to the supersymmetry algebra.

It is interesting to find (describe) all such deformations. We intend to  split this problem into two sub problems:

{\bf A} Find all infinitesimal deformations. By this we mean to solve the above problem up to order two in power series in ${\alpha'}$. Thus we should  look for ${\cal L}'(\nabla,\xi)={\cal L}^1(\nabla,\xi)$ and $\theta'_{\alpha }=\theta^1_{\alpha }$ such that 
\begin{equation}\label{E:dsfdjh}
\theta'_{\alpha }{\cal L}(\nabla,\xi)+\theta_{\alpha }{\cal L}(\nabla,\xi)'=\mbox{ full derivative }
\end{equation}
We may say that we are looking for on shell invariants of supersymmetry algebra.

We must also take into account a condition of on shell closure mod ${\alpha'}^2$ of  algebra of $\theta_{\alpha }({\alpha'})$. 

{\bf B} Extend an infinitesimal deformation to an actual deformation.

In this paper  we shall address  sub problem {\bf A}.

The gauge groups in our setup is $U(N)$, where $N$ is large.
Here is a list of the main results:

{\bf 1} A system of equations (\ref{E:dsfdjh}) is overdetermined. It is not clear apriory that it has nonzero solutions at all. It seems possible however to give a formula for a general solution, which is $Spin(10)$-invariant. It turns out that it is a sum of four terms 
\begin{equation}\label{E:fvwgb}
{\cal L}'(\nabla,\xi)={\cal L}'_{I}(\nabla,\xi)+{\cal L}'_{II}(\nabla,\xi)+{\cal L}'_{III}(\nabla,\xi)+{\cal L}'_{IV}(\nabla,\xi)
\end{equation}
\begin{equation}\label{E:dvsdjvh}
\begin{split}
&{\cal L}'_{I}(\nabla,\xi)=const_{I} tr\Bigg(  \frac{1}{8} F_{mn}F_{nr}F_{rs}F_{sm}- \frac{1}{32} \left( F_{mn}F_{mn}\right)^2\\ 
&+i\frac{1}{4} \xi^{\alpha} \Gamma_{m\alpha \beta} ({\nabla}_n\xi^{\beta}) F_{mr}F_{rn}  \\
&-i\frac{1}{8} \xi^{\alpha} \Gamma_{mnr\alpha \beta} ({\nabla}_s\xi^{\beta}) F^{mn} F_{rs}\\
&+ \frac{1}{8} \xi^{\alpha} \Gamma^m_{\alpha \beta} ({\nabla}_n\xi^{\beta})\xi^{\gamma} \Gamma_{m\gamma\delta} ({\nabla}_n\xi^{\delta}) \\
&- \frac{1}{4} \xi^{\alpha} \Gamma^m_{\alpha\beta} ({\nabla}_n\xi^{\beta})\xi^{\gamma} \Gamma_{n\gamma\delta} ({\nabla}_m\xi^{\delta}) \Bigg)\\
&\Bigg(\mbox{The formula for ${\cal L}'_{I}(\nabla,\xi)$ was borrowed from \cite{Berg}}\Bigg)\\
&{\cal L}'_{II}(\nabla,\xi)=const_{II}{\cal L}_3(\nabla,\xi)\quad \\
&\Bigg(\mbox{ formula 3.1 page 6 in \cite{Collinucci} is too long to be presented here}\Bigg)
\end{split}
\end{equation}
It is  $\alpha^{'2}$ coefficient of power series  (2.63) in \cite{Berg} .

At the moment we have only partial information about ${\cal L}'_{III}(\nabla,\xi)$ that it  one of two $\susy$ invariants of degree 5 in $\alpha'$. See section (\ref{S:ddgwshcx}) however.

Suppose $n(\nabla,\xi)$ is a noncommutative polynomial in covariant derivatives of curvature and spinor field. Define $\nu(\nabla,\xi)=trn(\nabla,\xi)$.

 Define 
\begin{equation}
{\cal L}'_{IV}(\nabla,\xi)=\epsilon^{\alpha_1\dots\alpha_{16}}\theta_{\alpha_1}\dots \theta_{\alpha_{16}}\nu(\nabla,\xi)
\end{equation}
 where $\epsilon$ is the Levi-Chevita tensor.

The theories with smaller supersymmetries which allow off shell formulation with manifest supersymmetries admit a simple construction of deformation $\L(\nabla,\xi)\rightarrow \L(\nabla,\xi)+\alpha'\L'(\nabla,\xi)$ of supersymmetric Lagrangian. $\L'(\nabla,\xi)$ can be chosen as an arbitrary superfunction in chiral superfields. In this case $\nu$ becomes one of the components of the chiral superfunction. The operator $\epsilon^{\alpha_1\dots\alpha_{16}}\theta_{\alpha_1}\dots \theta_{\alpha_{16}}$ can be interpreted as "chiral odd integration" (see \cite{DF} chapter 10). The analogy with theories with low number of supersymmetries is not accidental. In the course of the proof we used pure spinor approach of Howe-Berkovits (its homological version), which could be considered as the best approximation to manifestly susy-covariant formulation of Yang-Mills theory .


{\bf 2} Collection $\theta'_{\alpha }$ uniquely determine ${\cal L}'(\nabla,\xi)$ and visa versa.

Since everything is covariant with respect to the group of translations we may assume that the coefficients of the Lagrangian ${\cal L}'$ are some constants.  We can associate a degree with every Lagrangian $\L'=tr n(\nabla,\xi)$by the rule $deg \nabla_i =2, deg \xi^{\alpha}=3$. The degree of all constants is zero. As a result $deg {\cal L}(\nabla,\xi)=8$ for ${\cal L}(\nabla,\xi)$ defined in \ref{E:YM}. The relation of the degree $deg$ with more conventional degree in $\alpha'$ which we denote by $deg_{\alpha'}$ is 
\begin{equation}\label{E:aaghhs}
deg_{\alpha'}=\frac{deg-8}{4}
\end{equation}

{\bf 3} One can form a generating function $\tilde a(t)=\sum_{k\geq 0}\tilde  a_k t^k$. The coefficients $a_k$ are dimensions of linear spaces spanned by infinitesimal Lagrangians ${\cal L}'$ of degree $deg=k$, which satisfy (\ref{E:dsfdjh}), defined up to a field redefinition. Then
\begin{equation}\label{E:osdodsda}
\tilde a(t)=t^8(a(t)-126t^{-2}-144t)
\end{equation}
The formula for $a(t)$ is given in (\ref{E:fvdvbhads}).

A typical  Lagrangian  ${\cal L}'$ (no condition (\ref{E:dsfdjh}) is necessary at this point) is defined for a gauge group  $U(N)$, because it involves products (of derivatives) of curvature, like in $tr(F_{ij}F_{jk}F_{kl}F_{li})$ in (\ref{E:dvsdjvh}). The last Lagrangian is not defined for exceptional groups because it requires an additional data - a representation in $U(N)$. A possible restriction on ${\cal L}'$ is that it is defined for any Lie algebra, i.e. it is of the form 
\begin{equation}\label{E:cbvduf}
\sum_r (m_{r,1}(\nabla,\xi),m_{r,2}(\nabla,\xi))
\end{equation} 

In the last formula $(.,.)$ is the invariant dot product on the Lie algebra of the gauge group $\mathfrak{g}$, $m_{r,1},m_{r,2}$ are commutators in covariant derivatives of curvature and spinor fields. We call  Lagrangians written in  (\ref{E:cbvduf}) the Lagrangians of the Lie type. In this setup we can form a generating function $\tilde l(t)$ in analogy with $\tilde a(t)$. Then
\begin{equation}\label{E:knhbjh}
\tilde l(t)=t^8(l(t)-144t)
\end{equation}

The formula for $l(t)$ is given in (\ref{E:fvdvbhads}).

{\bf 4}
One can define a series of types of Lagrangians that are similar to the Lie type. They can be defined by the formula 
\begin{equation}
\L'(\nabla,\xi)=\sum_{\sigma\in S_p}\sum_{r} \pm tr(m_{r, {\sigma(1)}}(\nabla,\xi)\dots m_{r, {\sigma(p)}}(\nabla,\xi))
\end{equation} 
The commutators $m_{r, i}(\nabla,\xi)$ $1\leq i \leq p$ are defined as before. The auxiliary degree of such Lagrangian is equal to $p$. One can form a generating function $l(t,u)=\sum_{k,p}l_{kp}t^ku^p$, with $l_{kp}=dim L_{kp}$-dimensions of space of Lagrangians of bidegree $k,p$ up to field redefinition and satisfying (\ref{E:dsfdjh}). We do not provide a formula for $l(t,u)$ (our technique allows to do that but the formula becomes messy). Instead we tabulated below the fist few coefficients of a related function $\tilde l^{Spin(10)}(t,u)$. The  coefficient of $l^{Spin(10)}(t,u)$ are dimension of $Spin(10)$-invariants in spaces $L_{kp}$. In fact it is convenient to use $deg_{\alpha'}$ instead of $deg$ at this point, because $$L^{Spin(10)}_{k,p}=0\quad k\nequiv \quad0 \mbox{ mod } 4$$. A transition  $deg\rightarrow deg_{\alpha'}$ in the degrees of generating function  manifests in a change of variables $l^{Spin(10)}(t^{\frac{1}{4}},u)t^{-2}=\tilde l^{Spin(10)}(t,u)$ ($t^{-2}$-factor  is for agreement with (\ref{E:aaghhs})). The coefficients $\tilde l_{kp}$ of $\tilde l^{Spin(10)}(t,u)$ are
 \begin{equation}\label{T:dkodd1}
\mbox{
\scriptsize{
$\begin{array}{|c|c|c|c|c|c|c|c|c|c}
 &1&2&3&4&5&6&7&8&k=deg_{\alpha'}\\\hline
2& & &1& &1&3&18&172&\dots\\\hline
3& & & & & & &13&281&\dots\\\hline
4& &1& & &1&2&20 &267&\dots\\\hline
5& & & & & & &1 &68 &\dots\\\hline
6& & & & & & &1 &17 &\dots\\\hline  
7& & & & & & & & &\dots\\\hline
p&\dots&\dots&\dots&\dots&\dots&\dots&\dots &\dots &\dots
\end{array}$
}
}
\end{equation}

Let we say a few words about one important feature of Yang-Mills theory which was not accommodated in the present framework.

The Lagrangian \ref{E:YM} admits additional "trivial" translational supersymmetries:
$\tilde\theta_{\alpha}\nabla_i=0$, $\tilde\theta_{\alpha}\chi^{\beta}=\delta_{\alpha}^{\beta}$. A systematic treatment of these we defer to future publications.

The methods of this paper are similar to \cite{MSch},\cite{MSch2}, \cite{susyBar}. The principal ingredient of this paper is a Lie algebra $YM$. One can think about it as of a universal solution of D=10, N=1 Yang-Mills equations.

It comes about  as follows: we replace fields by generators, equations of motion by relations. In case of Yang-Mills equations this way we get $YM$ algebra. The supersymmetry generators also can be lifted to some formal variables. They act on $YM$ mimicking formulas of component formalism (\ref{E:dfagdsh}). We denote by $L$  the Lie algebra of derivations they generate.  This algebra was introduced in \cite{MSch} and studied in \cite{susyBar}. It turns out that $L$ contains $YM$ as an ideal. There is a projection $L\rightarrow \susy$ with a kernel $TYM$. It turns out that there is a tower of inclusions $TYM\subset YM\subset L$.

Suppose that we constructed some supersymmetric deformation   of D=10, N=1 Yang-Mills theory. The outlined above procedure will provide us with deformations $YM_{\alpha'}$ and $L_{\alpha'}$. Our main assumption is that in process of deformation dimensions of invariantly defined space do not jump. In particular a sequence of ideals $TYM_{\alpha'}\subset YM_{\alpha'}\subset L_{\alpha'}$   survives in the  process of deformation. 

Such algebraization allows us to use powerful methods of deformation theory of algebraic systems (Lie and associative algebras).

The paper is organized as follows:
We collected all necessary notations and definition   in sections (\ref{S:gadvb}), \ref{S:MD}.

In section (\ref{S:ibdhx}) we setup a stage. We define procisely how we understand a supersymmetric deformation of Yang-Mills theory . We have several flavors of deformations that are classified by types. We distinguish two types: associative ($\mathbf{A}$) and  Lie ($\mathbf{L}$) types. We also introduce an independent $\mathbf{Lg}$ condition  - the deformed Yang-Mills theory should have a Lagrangian.
We also introduce deformation complexes.

In section (\ref{S:hgdxsw}) we make the main reduction in deformation complexes, replacing the standard one by much smaller. It enables us to do the computations. We use pure spinors in essential way. The most important technical proposition in this section is (\ref{P:azbush}).

In  section \ref{SS:dvausdu} we give partial justification  of the claim {\bf 1} from the introduction .

In  section (\ref{S:gen})  we justify claims {\bf 3,4}.

In  section (\ref{S:xaadbr}) we verify claim {\bf 2} and finish with {\bf 1}.

We also provide an extensive appendix where some algebraic facts are collected and proofs of some technical statements are outlined.

In particular, we have a section \ref{S:Lquad} on quadratic algebras where the reader can find  explanations of  the main reduction from section (\ref{S:hgdxsw}).

We also have a section (\ref{S:gqagbj}) on equivariant cyclic homology. Some facts collected in it are useful for better understanding material from section \ref{SS:dvausdu}, (\ref{S:gen}).

\begin{acknowledgement}
The  author would like to thank  MPI, KITP, IAS, where the most of the work has been done. He also would like to thank  N.Nekrasov, M. Rocek, A.S.Schwarz,  D.Sullivan,  for useful discussions and comments . Also I would like to thank B. Lowson and   M.Rocek for opportunity to present some of this material at "2005 Simons workshop".
\end{acknowledgement}

\subsection{Notations}\label{S:gadvb}
All linear spaces in this note are defined over complex numbers $\mathbb{C}$.

An abbreviation for a complex $$\dots \rightarrow A_i\rightarrow A_{i+1}\rightarrow \dots $$ is $A^{\bullet}$. There is a standard shift operation on complexes $(A^{\bullet}[n])_{i}=(A^{\bullet})_{i+n}$, $d_{A^{\bullet}[n]}=(-1)^nd_{A^{\bullet}}$.

Let $C$ be an algebra, $\varepsilon:C\rightarrow \mathbb{C}$ a homomorphism (augmentation). Denote $IC=\Ker \varepsilon$

We denote $\Sym(W)=\bigoplus_{i\geq 0}\Sym^i(W)$ a symmetric algebra of a linear space $W$. Denote $a\bullet b$ a product of elements $a,b\in \Sym(W)$. The object $\Lambda(W)=\bigoplus_{i\geq 0}\Lambda^i(W)$ is the Grassman algebra on $W$. If $W$ is a graded vector space that $\Sym(W),\Lambda(W)$ are defined conforming to the sign rules.

We denote a bracket in a Lie algebra by $[.,.]$ or $\{.,.\}$. We shall use uniform notations for commutators and anticommutators.

We use Einstein convention of summation over repeated indices and  do not use $\sum$ sign. In case when $\sum$ is present we perform a summation over non-tensor  indices, the range of summation can be guessed from the context.

Let $V$ be $10$-dimensional linear space over complex numbers. It is  equipped with a symmetric nondegenerate dot-product invariant with respect to $Spin(10)$. Let $S$ be an irreducible complex spinor representation of $Spin(10)$. Vectors $v_1,\dots,v_{10}$ is an orthonormal basis of $V$, $\chi^{\alpha}$ $\alpha=1,\dots, 16$ is a basis of $S$. $\theta_{\alpha}$ is a dual basis of $S^*$.

This data allows us define $\Gamma$-matrices $\Gamma_{\alpha\beta}^i$, $\Gamma^{\alpha\beta}_i$,  $\Gamma_{\alpha}^{\beta ij}$, $\Gamma_{\alpha\beta}^{i_1i_2i_3} ,\dots \Gamma^{\alpha\beta}_{i_1i_2i_3i_4i_5}\dots$. See \cite{susyBar} for discussion of spinors and $\Gamma$-matrices.

All tensors involved in our considerations are build from vector representation $V$ (latin vector indices), irreducible spinor representations $S,S^*$ (Greek spinor indices) of $Spin(10)$. A presence of a dot-product  permits us to make no distinction between lower and upper vector indices. We however shall make a careful distinction between lower and upper spinor indices. 

Denote $f\circ g$ a composition of maps: $(f\circ g)(x)=f(g(x))$.

We use the standard convention labeling representation of a semisimple group  by it highest weight.
The Dynkin graph of the group $Spin(10)$ is
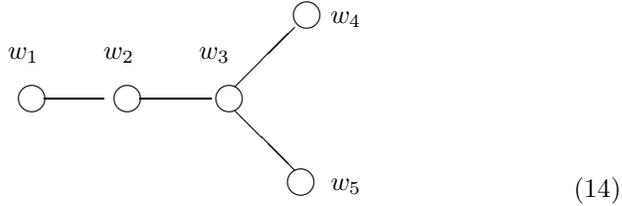
\begin{equation}\label{P:kxccvg}
\mbox{
\setlength{\unitlength}{3947sp}
\begin{picture}(2025,1246)(301,-452)
\thinlines
\put(1688,164){\circle{168}}
\put(451,164){\circle{168}}
\put(1051,164){\circle{168}}
\put(2176,689){\circle{168}}
\put(2138,-361){\circle{168}}
\put(1126,164){\line( 1, 0){450}}
\put(1726,239){\line( 1, 1){375}}
\put(1726, 89){\line( 1,-1){375}}
\put(526,164){\line( 1, 0){375}}
\put(301,389){\makebox(0,0)[lb]{$w_1$}}
\put(901,389){\makebox(0,0)[lb]{$w_2$}}
\put(1501,389){\makebox(0,0)[lb]{$w_3$}}
\put(2326,-436){\makebox(0,0)[lb]{$w_5$}}
\put(2326,614){\makebox(0,0)[lb]{$w_4$}}
\end{picture}
}
\end{equation}
The labels $w_i$ correspond to coordinates of the highest weight.
We  encode a representation that is labeled by Dynkin diagram with labels as above by an array $[w_1,w_2,w_3,w_4,w_5]$.
Our convention is that spinor representation $S_1\subset S$ is equal to
irreducible representation with highest weight $[0,0,0,0,1]$, the
tautological representation in $\mathbb{ C}^{10}$ is equal to
$[1,0,0,0,0]$, the adjoint representation is equal to
$[0,1,0,0,0]$... A general representation is then $\bigoplus_{w_i\geq 0}a_{w_1,\dots w_5}[w_1,\dots,w_5]$, $a_{w_1,\dots w_5}\in \mathbb{Z}_{\geq 0} $ are the multiplicities.

\subsection{Main definitions.}\label{S:MD}
\begin{definition}\label{D:qxdt}
$YM$ algebra is a quotient of a free graded Lie algebra $\widetilde{YM}=Free<v_1,\dots,v_n,\chi^{1},\dots,\chi^{16}>$, $deg(v_i)=2, deg(\chi^{\alpha})=3$ by an ideal. The ideal is generated by relations
\begin{align}
&\tilde v_m=[v_{s},[v_{s},v_{m}]]-\frac{1}{2}\Gamma_{\alpha \beta}^m[\chi^{\alpha},\chi^{\beta}]\label{E:jhda} \\
&\tilde \chi_{\alpha}=\Gamma_{\alpha \beta}^s[v_s,\chi^{\beta}]\label{E:jhda2}
\end{align}
\end{definition}

\begin{definition}
A Lie algebra $L$ is a quotient of a free graded Lie algebra $Free<\theta_1,\dots,\theta_{16}>$. $deg\theta_{i}=1$. The generators of the ideal of relations are 
\begin{equation}\label{E:iagxdqdd}
\Gamma^{\alpha \beta}_{i_1\dots i_5}[\theta_{\alpha},\theta_{\beta}]=0
\end{equation}
\end{definition}

From \cite{MSch}, \cite{MSch2} we know that $YM$ is a subalgebra of Lie algebra $L$.
The embedding $\rho:YM\rightarrow L$ is defined by the formulas
\begin{equation}
\begin{split}
&\rho(v_i)=\Gamma^{\alpha\beta}_i[\theta_{\alpha},\theta_{\beta}]\\
&\rho(\chi^{\alpha})=\Gamma^{\alpha\beta s}[\rho(v_s),\theta_{\beta}]
\end{split}
\end{equation}

Denote 
\begin{equation}
L=\bigoplus_{i\geq 1} L_i
\end{equation}
 -decomposition into graded pieces. It was proved in \cite{susyBar} that 
\begin{equation}
YM=\bigoplus_{i\geq 2} L_i
\end{equation}
In the following we identify $YM$ with subalgebra of $L$ such that the following formula holds $[\theta_{\alpha},\theta_{\beta}]=\Gamma_{\alpha\beta}^iv_i$

The algebra $YM$ is an ideal in $L$. We interpret the adjoint action of $\theta_{\alpha}\in L_1$ on $YM$ as an action of supersymmetries: 
\begin{equation}\label{E:dfagdsh}
\begin{split}
&\theta_{\alpha}v_i=\Gamma_{\alpha\beta i}\chi^{\beta} \\
&\theta_{\beta}\chi^{\alpha}=\Gamma^{\alpha ij}_{\beta }F_{ij}
\end{split}
\end{equation}

 Following \cite{MSch2} and \cite{susyBar} denote 
\begin{equation}\label{E:cxvfbw}
TYM=\bigoplus_{i\geq 3} L_i
\end{equation}

From \cite{MSch2} we know that $TYM$ is generated as a Lie algebra by elements of the form 
\begin{equation}\label{E:dvuvg}
\begin{split}
&[v_{i_{1}},\dots,[v_{i_{k}},F_{ij}]\dots] \quad F_{ij}=[v_i,v_j]\\
&[v_{i_{1}},\dots,[v_{i_{k}},\chi^{\alpha}]\dots]
\end{split}
\end{equation}
\begin{definition}
Denote by $\susy$ a graded variant of Lie algebra of supersymmetries in dimension ten.  It is $\mathbb{Z}$-graded algebra $\susy_1=S^*=<\theta_1,\dots,\theta_{16}>, \susy_2=V=<v_1,\dots,v_{10}>$. After reduction of grading modulo two we get a standard $\mathbb{Z}_2$ grading on  $\susy$. The space $\susy_2$ is the center. The commutator of $\theta_{\alpha}\in \susy_1$ is defined  by the formula $[\theta_{\alpha},\theta_{\beta}]=2\Gamma_{\alpha\beta}^iv_i$. 
\end{definition}
%
%
%

Let us remind the standard relation of $YM$-algebra to the classical Yang-Mills theory.

Let $\nabla$ be a connection in a principle $U(N)$-bundle on $\mathbb{R}^{10}$. We assume that a choice of coordinates $x_i$ on $\mathbb{R}^{10}$ is given in which a metric has a diagonal form $dx_i^2$.
Denote $S^*\otimes\mathfrak{u}(N)$- a tensor product of irreducible complex dual spinor and adjoint bundles. Let $\tilde\xi_{\alpha}$, $\alpha=1,\dots 16$ be a set of sections which form a basis of $S$ (pointvise). Then any section of $S\otimes\mathfrak{u}(N)$ can be presented as $\tilde\xi_{\alpha} \xi^{\alpha}$, where $\xi^{\alpha}$ is sixteen $\mathfrak{u}(N)$-valued functions.
Denote $\nabla_i=\nabla_{\frac{\partial \ \ }{\partial x_i}}$. Let us assume that the $\Gamma$-matrices $\Gamma^{i}_{\alpha\beta}$ in the bases are translationary invariant and equal to the standard $\Gamma$-matrices used in definition (\ref{D:qxdt}).

Connection in the principal bundle define covariant derivative in any associated bundle, which we denote by the same letter $\nabla$. We shall be interested in $\Lambda(\Theta)\otimes T$, where $T$ is the fundamental representation of $U(N)$.
We can interpret $\Lambda(\Theta)$ as function on auxiliary odd space of parameters $\Theta^*$. As it is common in supergeometry  connection $\nabla$ has coefficients in $\Lambda(\Theta)$. The same applies to spinor: $\xi^{\alpha}$ is a matrix spinor with coefficients in $\Lambda(\Theta)$, i.e. section of $S^*\otimes \Lambda(\Theta)\otimes \mathfrak{u}(N)$.

Levi-Chevita connection, together with connection $\nabla$  define covariant derivatives $\nabla_i$ acting in sections of $\Lambda(\Theta)\otimes T$, which we can consider  as differential operators of the first order. Any odd section of  $\Lambda(\Theta)\otimes\mathfrak{u}(N)$ defines a matrix operator with $\Lambda(\Theta)$ entries acting on $\Lambda(\Theta)\otimes T$. For a given choice of connection $\nabla_i$ and spinors $\xi^{\alpha}$ they generate a subalgebra in a (super)Lie algebra of differential operators acting in  $\Lambda(\Theta)\otimes T$.

An  assignment 
\begin{equation}\label{E:qdscdfre}
\begin{split}
&v_i\rightarrow \nabla_i\\
&\chi^{\alpha}\rightarrow \xi^{\alpha}
\end{split}
\end{equation} is a homomorphism of $YM$-algebra to the algebra of differential operators if and only if $\nabla_i, \xi^{\alpha}$ is a solution of classical Yang-Mills equation
\begin{equation}
\begin{split}
&\nabla_iF_{ij}=\frac{1}{2}\Gamma^j_{\alpha\beta}[\xi^{\alpha},\xi^{\beta}], \quad [\nabla_i,\nabla_j]=F_{ij}\\
&\Gamma^i_{\alpha\beta}\nabla_i\xi^{\beta}=0
\end{split}
\end{equation}
Standard supersymmetry transformations can be obtained from (\ref{E:dfagdsh}) after a substitution (\ref{E:qdscdfre})
 These transformations satisfy relation (\ref{E:iagxdqdd}) if $\nabla_i, \xi^{\alpha}$ is a solution of classical Yang-Mills equation.

\section{Definition of supersymmetric deformation of YM algebra}\label{S:ibdhx}

Fix Lie algebras $\mathfrak{n}$ and $\mathfrak{l}$.
\begin{definition}
A Lie algebra $\mathfrak{g}$ belongs to the class $\mathbf{L}(\mathfrak{n},\mathfrak{l})$ if $\mathfrak{g}$ is an extension of $\mathfrak{n}$ by $\mathfrak{l}$,i.e fits into exact sequence $$0 \rightarrow \mathfrak{l}\rightarrow \mathfrak{g}\rightarrow\mathfrak{n}\rightarrow 0$$ .
\end{definition}

By construction we have $\susy=L/TYM$. It means that $L\in \mathbf{L}(\susy,TYM)$

In the following text ${\alpha'}$ is a formal parameter of deformation. All maps involved in the construction are $\mathbb{C}[[{\alpha'}]]$-linear. All deformations are flat modules over $\mathbb{C}[[{\alpha'}]]$. We have an isomorphism of $\mathbb{C}[[{\alpha'}]]$ modules $L_{\alpha'}=L\hat{\otimes}\mathbb{C}[[{\alpha'}]]$. Lie algebra has a topology defined by filtration $F^i=\bigoplus_{k\geq i}L_k$ (the reader can consult  \cite{MSch2} about completions).

To shorten the notations we denote a trivial deformation of Lie algebra $\mathfrak{g}$ as $\mathfrak{g}({\alpha'})$. It is isomorphic to $\mathfrak{g}\hat{\otimes}\mathbb{C}[[{\alpha'}]]$ not just as a $\mathbb{C}[[{\alpha'}]]$-module but as a Lie algebra.
\begin{definition}
We shall be interested in deformations $L_{{\alpha'}}$ of $L$ which contains a Lie subalgebra $TYM_{\alpha'}$ such that there is a short exact sequence of algebras
\begin{equation}\label{E:cond}
0\rightarrow TYM_{\alpha'} \rightarrow L_{{\alpha'}} \overset{p}{\rightarrow} \susy({\alpha'}) \rightarrow 0
\end{equation} 
and $\susy$ stays undeformed. Two deformations $L_{\alpha'}$ and $L_{\alpha'}'$are equivalent if there is an isomorphism $\eta:L_{\alpha'}\rightarrow L_{\alpha'}'$, which map $TYM_{\alpha'}$ into  $TYM_{\alpha'}$. We shall call such class of deformations- $\mathbf{L}(\susy({\alpha'}),TYM_{\alpha'})$-deformations.
\end{definition}
\begin{remark}
The algebra $TYM$ is free (see \cite{MSch2}), therefore are rigid . The class $\mathbf{L}(\susy({\alpha'}),TYM_{\alpha'})$ coincides with $\mathbf{L}(\susy({\alpha'}),TYM({\alpha'}))$. 
\end{remark}
%
%
%
\begin{definition}
 $\mathbf{L}^{Spin(10)}(\susy({\alpha'}),TYM({\alpha'}))$ a subclass of $\mathbf{L}(\susy({\alpha'}),TYM({\alpha'}))$ of $Spin(10)$-equivariant  deformation . The $Spin(10)$ representation content of $L_{\alpha}$ must coincide with the content of $L$.
\end{definition}
A deformation of $L$ of type $\mathbf{L}(\susy({\alpha'}),TYM({\alpha'}))$ gives rise to a deformation of $YM$ algebra. Indeed we can take $YM_{\alpha'}=[L_{\alpha'},L_{\alpha'}]=\{\sum_i[a_i,b_i]|a_i,b_i\in L_{\alpha'}\}$. $YM_{\alpha'}$ is an ideal in $L_{\alpha'}$. We recover the action of supersymmetries from adjoint action of generators of $L_{\alpha'}$ on $YM_{\alpha'}$. 

An algebra  $L_{\alpha'}\in \mathbf{L}(\susy({\alpha'}),TYM({\alpha'}))$ admits a decreasing filtration $F^i=[L_{\alpha'},\dots,[L_{\alpha'},L_{\alpha'}]\dots]-i$ times. Except few possible pathological cases $\bigcap_iF^i=0$, the last condition holds true if $L_{\alpha'}\in \mathbf{L}^{Spin(10)}(\susy({\alpha'}),TYM({\alpha'}))$.

The formal definition of deformation  can be  translated into the language of connections. 
The number of generators and relations of $YM_{\alpha'}$ is the same as of $YM$, if $L_{\alpha'}\in \mathbf{L}^{Spin(10)}(\susy({\alpha'}),TYM({\alpha'}))$. This is because of the mentioned filtration. On $YM$ the filtration is equal to $F^i=\bigoplus _{k\geq i}YM_k$. We identify $YM_{\alpha'}$ with $YM\otimes \mathbb{C}[[\alpha']]$ as $\mathbb{C}[[\alpha']]$-modules via identity transformations. This enables us to interpret algebraic generators of $YM$ as  generators of $YM_{\alpha'}$ over  $\mathbb{C}[[\alpha']]$
%
 The deformed relations must then be of the form 
\begin{equation}\label{E:idfhvqe}
\begin{split}
&[v_i[v_i,v_j]]-\frac{1}{2}\Gamma^j_{\alpha\beta}[\chi^{\alpha},\chi^{\beta}]={\alpha'}r_j({\alpha'})\\
&\Gamma^i_{\alpha\beta}[v_i,\chi^{\beta}]={\alpha'}s_{\alpha}({\alpha'})
\end{split}
\end{equation}
 The elements $r_j({\alpha'}), s_{\alpha}({\alpha'})$ are some formal power series with coefficients in ideal of free Lie algebra $\widetilde{YM}$ generated by (\ref{E:dvuvg}). After substitution (\ref{E:qdscdfre}) the remainders $r_j({\alpha'}),s_{\alpha}({\alpha'})$ will become formal powers series with coefficients in Lie algebra, generated by elements (\ref{E:idhev})
\begin{equation}\label{E:idhev}
\begin{split}
&\nabla_{i_1}\dots\nabla_{i_k}F_{st} \\
&\nabla_{i_1}\dots\nabla_{i_k}\xi^{\alpha}
\end{split}
\end{equation} 
 for $k\geq 0$.

We say that $\nabla_i,\xi^{\alpha}$ is a solution of the deformed Yang-Mills  equation if  $\nabla_i,\xi^{\alpha}$ to satisfy (\ref{E:idfhvqe}) after substitution (\ref{E:qdscdfre}) .


There is a standard language that was proved to be useful in solving deformation problems of above type. It is the language of deformation theory of Lie algebras

\begin{definition}\label{D:qhashdnf}
Suppose  $\mathfrak{g}$ is an arbitrary Lie algebra and $N$ is a $\mathfrak{g}$-module. It is a homomorphism $\rho:\mathfrak{g}\rightarrow End(N)$. There is a complex 
\begin{equation}\label{E:xhafq}
C^k( \mathfrak{g}, N)=Hom(\Lambda^k(\mathfrak{g}),N)
\end{equation}
, called Cartan-Chevalley complex. The differential $d:C^k( \mathfrak{g},N)\rightarrow C^{k+1}( \mathfrak{g},N)$ is defined by the formula:
\begin{equation}\label{E:diff}
\begin{split}
&(dc)(l_1,\dots,l_{k+1})=\sum_{i=1}^{k+1}(-1)^i\rho(l_i)c(l_1,\dots,\hat l_i,\dots,l_{k+1})+\\
&+\sum_{i<j}(-1)^{i+j-1}c([l_i,l_j],l_1,\dots,\hat l_i,\dots,\hat l_j,\dots,l_{k+1})
\end{split}
\end{equation}
The cohomology of this complex is denoted by $H^k(\mathfrak{g},N)$. 

A complex
\begin{equation}\label{E:xhafq12}
C_k( \mathfrak{g}, N)=\Lambda^k(\mathfrak{g})\otimes N
\end{equation}
has a differential 
\begin{equation}\label{E:diffh}
\begin{split}
&d:C_k( \mathfrak{g},N)\rightarrow C_{k-1}( \mathfrak{g},N)\\
&d (n\otimes l_1\wedge\dots\wedge l_{k})=\sum_{i=1}^{k}(-1)^i\rho(l_i)n\otimes l_1 \wedge \dots \wedge \hat l_i \wedge \dots \wedge l_{k}+\\
&+\sum_{i<j}(-1)^{i+j-1} n\otimes [l_i,l_j]\wedge \dots \wedge \hat l_i\wedge \dots \wedge \hat l_j \wedge \dots \wedge l_{k}\quad l_s\in \mathfrak{g},n\in N
\end{split}
\end{equation}
The cohomology of this complex is denoted by $H_k(\mathfrak{g},N)$. There is an obvious extension of these constructions to $\mathbb{Z}_2$ or  $\mathbb{Z}$ graded category. For details the reader can consult \cite{Fuchs} .
\end{definition}

For deformation purposes we shall be interested in adjoint representation $N=\mathfrak{g}$ or adjoint representation in universal enveloping $U(\mathfrak{g})$.
We shall give some illustrations of usefulness of complex $C^{\bullet}( \mathfrak{g}, N)$. For simplicity in the introductory discussion we assume that $\mathfrak{g}$ is a purely even Lie algebra.

Let us start with $c\in C^1( \mathfrak{g}, \mathfrak{g})$. It is easy to see that the condition $dc=0$ is the equation $c([l_1,l_2])=[l_1,c(l_2)]+[c(l_1),l_2]$. It is a condition that $c$ is a derivation of $\mathfrak{g}$: $c\in Der(\mathfrak{g})$ . There is a class of trivial derivations $In(\mathfrak{g})$-so called inner derivations $c_a(l)=[a,l]$. It is natural to work with a quotient 
\begin{equation}\label{E:dsfhgd}
Der(\mathfrak{g})/In(\mathfrak{g})=Out( \mathfrak{g})
\end{equation} 
The later linear space by definition coincides with $H^1(\mathfrak{g},\mathfrak{g})$. 

As an exercise the reader can check that $H^0(\mathfrak{g},\mathfrak{g})$ coincides with the center of $\mathfrak{g}$.

Any derivation of  $\mathfrak{g}$ defines a derivation of universal enveloping  $U(\mathfrak{g})$. The converse is not true. If one would like to understand derivations of $U(\mathfrak{g})$, one has to replace $N$ by $U(\mathfrak{g})$ in (\ref{E:xhafq}) and compute the first cohomology.

The groups  $H^k(\mathfrak{g},\mathfrak{g})$  are a direct summands in  $H^k(\mathfrak{g},U(\mathfrak{g}))$.

 The linear space $H^2(\mathfrak{g},\mathfrak{g})$ can be interpreted as a space of nonequivalent infinitesimal deformations of  $\mathfrak{g}$.
Indeed if we expand a deformed bracket $[.,.]_{\alpha'}$ into formal series in ${\alpha'}$, we shall get 
\begin{equation}
[.,.]_{\alpha'}=[.,.]+{\alpha'}\gamma_1(.,.)+{\alpha'}^2\gamma_2(.,.)+\dots
\end{equation}

Denote $\gamma(.,.)=\gamma_1(.,.)$. The map $a_1\otimes a_2 \otimes a_3 \rightarrow \sum_{\sigma \in \mathbb{Z}_3\subset S_3}[[a_{\sigma(1)},a_{\sigma(2)}]_{\alpha'},a_{\sigma(3)}]_{\alpha'}$ (which is zero for a Lie algebra) has it first Taylor coefficient equal to
\begin{equation}
\sum_{\sigma \in \mathbb{Z}_3\subset S_3}[\gamma(a_{\sigma(1)},a_{\sigma(2)}),a_{\sigma(3)}] +\gamma([a_{\sigma(1)},a_{\sigma(2)}],a_{\sigma(3)})
\end{equation}
It is equal to zero precisely when $\gamma$ is two-cocycle. Denote a space of two cocycles by $Z^2$. There is a subspace $B^2\subset C^2(\mathfrak{g},\mathfrak{g})$, which is generated by infinitesimal action of Lie algebra of coordinate change $\mathfrak{gl}(\mathfrak{g})$ on the bracket, viewed as an element $e\in C^2(\mathfrak{g},\mathfrak{g})$. We may identify $\mathfrak{gl}(\mathfrak{g})$ with  $C^1(\mathfrak{g},\mathfrak{g})$. Denote the action of $l\in \mathfrak{gl}(\mathfrak{g})$ on $e$ by $le$. Then $le=d(l)$. We see that $dB^2=0$. As a result a tangent space to the space of deformations of $\mathfrak{g}$ is equal to  $Z^2/B^2=H^2(\mathfrak{g},\mathfrak{g})$. 
 Recalling   discussion of derivations the reader should not be surprised to know that  the space  $H^2(\mathfrak{g},U(\mathfrak{g}))$ classifies infinitesimal deformations of $U(\mathfrak{g})$.


 Deformation of the bracket $[a,b]_{\alpha'}=[a,b]+{\alpha'}\gamma(a,b)+\dots $ of $L$ which satisfies (\ref{E:cond}) must satisfy $p\gamma=0$. Hence $\Im \gamma \subset TYM$. The following proposition becomes  obvious 
\begin{proposition}
 Infinitesimal deformations of $L$ which can be put into short exact sequence (\ref{E:cond}) are classified by elements of $H^2(L,TYM)$, infinitesimal deformations $\mathbf{L}^{Spin(10)}(\susy({\alpha'}),TYM({\alpha'}))$ are classified  $H^2(L,TYM)^{Spin(10)}$. In both cases, the degrees of deformations are even.
\end{proposition}
The elements of $H^i(L,TYM)$ are graded. The grading of $TYM$ starts with three. For a homogeneous cocycle $\gamma(a,b)\in C^2(L,TYM)$ its degree can be determined by a formula $deg(\gamma(\theta_{\alpha},\theta_{\beta}))-2$. It is grater then zero. An additional restriction on the cocycle is that its degree must be even.

It is also useful to study cohomology of $H^2(L,U(TYM))$. We provide below a somewhat cumbersome definition of deformation of algebra $U(L)$ which leads to the space of infinitesimal deformations equal to $H^2(L,U(TYM))$. A difficulty is that $U(L)_{\alpha'}$ is no longer a universal enveloping. It however retains certain properties of universal enveloping , which enables us to interpret the action of $\theta_{\alpha}(\alpha')$ by commutators on $U(TYM)$ as supersymmetries.

One can define a Lie algebra $Out(C)$ for an associative algebra $C$. The definition mimics to the Lie algebra case (\ref{E:dsfhgd}).

The linear space $L_1+L_2\subset L$ normalizes Lie subalgebra $TYM$. It also normalizes $U(TYM)\subset U(L)$. This data defined a homomorphism $\susy\rightarrow Out(U(TYM))$. 

An abstraction of the above observation is a homomorphism 
\begin{equation}\label{E:dfdsfh}
h:\mathfrak{n}\rightarrow Out(C)
\end{equation}
,  where $\mathfrak{n}$ is a (graded) Lie algebra ,  $C$ is some (graded) associative algebra. 

 Under some assumptions, we can construct algebra $B$ which enjoys  the following list of properties:

{\bf 1} There an isomorphism of linear spaces
\begin{equation}\label{E:isom}
\Sym(\mathfrak{n})\otimes C\overset{\mu}{\cong} B
\end{equation}, 

{\bf 2} $C$ is a subalgebra of $B$, $\mu$ restricted on $C$ is a homomorphism.

{\bf 3} For  $l,l_1,l_2\in \mathfrak{n}$ $[\mu(l),\mu(C)]\subset \mu(C)$, $[\mu(l_1),\mu(l_2)]=\mu([l_1,l_2])+\gamma(l_1,l_2)\subset \mu(\mathfrak{n})+C$. We require that it defines  a homomorphism $\mathfrak{n}\rightarrow Out(C)$.

{\bf 3'} In case when $C$ has an augmentation (as in case of universal enveloping ) $\gamma(l_1,l_2)\in IC, [\mu(l),\mu(IC)]\subset \mu(IC)$

{\bf 4} A map 
\begin{equation}
l_1\bullet\dots\bullet l_n\otimes c \rightarrow \frac{1}{n!}\sum_{\sigma\in S_n}\pm \mu(l_{\sigma(1)})\dots \mu(l_{\sigma(n)})\mu(c)
\end{equation}
induces the isomorphism (\ref{E:isom}).
\begin{definition}
We say that the algebra belongs to the class $\mathbf{A}(\mathfrak{n},C)$ ($\mathbf{A}'(\mathfrak{n},C)$) if it satisfies assumptions {\bf 1,2,3,4}( {\bf 1,2,3',4}).
\end{definition}

We provide a brief sketch of construction of $B\in \mathbf{A}(\mathfrak{n},C)$ given a homomorphism $h$  ((\ref{E:dfdsfh})).

The map $h$ can be used  to construct  an extension of Lie algebras
\begin{equation}\label{E:extwqw}
0 \rightarrow In(C) \rightarrow \tilde{\mathfrak{g}} \rightarrow \mathfrak{n} \rightarrow 0
\end{equation}
$In(C)=C/Z(C)$; $Z(C)$ is the center of $C$. 

Denote by $L(C)$  a Lie algebra with linear space $C$ and bracket defined by commutator in the associative algebra $C$.

In general it is not possible to lift  (\ref{E:extwqw}) to an extension 
\begin{equation}\label{E:qcsdoi}
0\rightarrow L(C) \rightarrow \mathfrak{g} \rightarrow \mathfrak{n} \rightarrow 0
\end{equation}
with an obstruction in $H^3(\mathfrak{n},Z(C))$.
\begin{assumption}
Let us assume that $Z(C)=\mathbb{C}$ and there is an augmentation $C\rightarrow \mathbb{C}$. Then $L(C)=In(C)+Z(C)$ as a Lie algebra. This is satisfied if $C$ is a free algebra. 

In such case we have a trivial lift of (\ref{E:extwqw}) to (\ref{E:qcsdoi}).
\end{assumption}
The universal enveloping $U(\mathfrak{g})$ contains $U(LC)$ . There is a canonical homomorphism of associative algebras $U(LC)\rightarrow C$. Define $B=U(\mathfrak{g})\underset{U(LC)}{\otimes }C$- a free product of algebras. The isomorphism $B\cong \Sym(\mathfrak{g})\otimes C$ comes from Poincare-Birkhoff-Witt theorem.

Our experience shows that the scope of deformations of $L$ in class \\  $\mathbf{L}(\susy(\alpha'),TYM(\alpha'))$  is a bit narrow . We expand it in the following
\begin{definition}
We say that a deformation $U(L)_{\alpha'}$ of $U(L)$ is of $\mathbf{A}$($\mathbf{A}'$)-type if it belongs to the class $\mathbf{A}(\susy({\alpha'}),U(TYM)({\alpha'}))$ ($\mathbf{A}'(\susy({\alpha'}),U(TYM)({\alpha'}))$ ).


\end{definition}
It is clear that any deformation of $\mathbf{L}$-type is of $\mathbf{A}$-type (take a universal enveloping algebra). The converse is not true.

In a study of deformations of type $\mathbf{A}(\susy({\alpha'}),U(TYM)({\alpha'}))$  the theory of deformations of associative algebras might seem more relevant .

Suppose we have a deformation $L_{\alpha'}\in\mathbf{A}(\susy({\alpha'}),U(TYM)({\alpha'}))$
 Such deformations are governed by Hochschild cohomology (see section \ref{S:Lquad}). There is a comparison result (\ref{R:wush}) which asserts 
that a deformation cocycle $\gamma(a,b)$ is completely determined by it values on $\Lambda^2(L)\subset U(L)\otimes U(L)$. 

If $a,b\in TYM$, then $\gamma(a,b)\in U(TYM)$, because $U(TYM)({\alpha'})$ is a subalgebra of $U(L)_{\alpha'}$\footnote{As in Lie algebra case $ U(TYM)$ is rigid}.

If $a\in L_1+L_2, b\in TYM$ then $\gamma(a,b)\in U(TYM)$, because $[a,U(TYM)({\alpha'})]_{\alpha'}\subset U(TYM)({\alpha'})$.

If $a,b\in L_1+L_2$ then $\gamma(a,b)\in U(TYM)$, because $L_1+L_2$ generates homomorphism of $\susy$ into $Out(U(TYM)({\alpha'}))$ and Lie algebra $\susy$ remains undeformed.

We summarize previous discussion in the following
\begin{proposition}
Infinitesimal deformations of algebra $U(L)$ in a class of algebras $\mathbf{A}(\susy({\alpha'}),U(TYM)({\alpha'}))$ ($\mathbf{A}'(\susy({\alpha'}),U(TYM)({\alpha'}))$) are parametrized by points of a linear space $H^2(L,U(TYM))$( $H^2(L,IU(TYM))$) of even degree. $H^2(L,U(TYM))^{Spin(10)}$( $H^2(L,IU(TYM))^{Spin(10)}$) corresponds to $Spin(10)$ equivariant infinitesimal deformation. 
\end{proposition}

The algebra $U(L)_{\alpha'}$ contains a subalgebra $U(YM)_{\alpha'}$ generated by $\mu(\susy_2)({\alpha'})$- even part of $\susy$ and $U(TYM)({\alpha'})$. The relations in $U(YM)_{\alpha'}$ have a form of (\ref{E:idfhvqe}), $r_j({\alpha'})$, $s_{\alpha}({\alpha'})$ have Taylor coefficients in associative algebra generated by (\ref{E:dvuvg}). If we take a realization of this theory by connections and spinors, the perturbed part of YM-equation is a product of covariant derivatives of curvature and spinor fields.

Let us back up  to the definition of deformations of type $\mathbf{L}(\susy({\alpha'}),TYM({\alpha'}))$.
In this setup the deformed Lie algebra $L_{\alpha'}$ contains a subalgebra $YM_{\alpha'}$.


In our dictionary, where generators of $YM$ correspond to the fields, cyclic words in generators correspond to Lagrangians. Indeed substituting fields for variables, taking trace multiplying on the volume form we get  a Lagrangian. For this  purpose is not necessary to use cyclic-ordinary words will do as good. However, representation of a Lagrangian as sum of ordinary words is redundant: trace has a cyclic symmetry as a result it picks up precisely cyclic words .

The algebraic analysis uses the following formalism. We have a short exact sequence
\begin{equation}
0\rightarrow I\rightarrow \widetilde{YM} \overset{p}{\rightarrow} V\rightarrow 0
\end{equation} 
The linear space $V$ is equipped with a zero bracket. The map $p$ projects $v_1,\dots,v_{10}$ into a basis of $V$. The ideal $I$ is generated by $F_{ij}=[v_i,v_j],\chi^{\alpha}$. As a Lie algebra it is  generated by symbols $[v_{(i_1},[\dots [v_{i_k)},F_{ij}]$, $[v_{(i_1},[\dots [v_{i_k)},\chi^{\alpha}]$, where $()$denote symmetrization , modulo relation $[v_i,F_{jk}]+[v_k,F_{ij}]+[v_j,F_{ki}]=0$ (Bianchi identity). Denote mentioned space of generators by $\widetilde{M}$. The Lie algebra $I$ is free. Making substitution (\ref{E:qdscdfre}) into an element $n$ we recover $n(\nabla,\xi)$ from the introduction. The universal enveloping $U(I)$ has the same set of generators as $I$. The space of cyclic words \footnote{Working in a graded setting we must take  a grading into account while defining cyclic word} in an alphabet defined by a basis of $\widetilde{M}$ is $Cyc(U(I))$ is equal to $U(I)/[U(I),U(I)]$. The commutator $[U(I),U(I)]$ as linear space is generated by $ab-(-1)^{deg(a)deg(b)}ba$.  The space $Cyc(U(I))$ has a homological interpretation. It is equal to $HH_0(U(I),U(I))=H_0(I,U(I))$ (see sections \ref{S:Lquad} for definition $HH_{\bullet}$). The Lie algebra $\widetilde{YM}$ acts on $Cyc(U(I))$ through abelian $V$. We have described above how to build a Lagrangian density $\L'(\nabla,\xi)$ out of element $\L'\in Cyc(U(I))$. It is clear that $(v_i\L')(\nabla,\xi)=\frac{\partial \L'(\nabla,\xi)}{\partial x_i}$. It means the action associated with $v_i\L'$ is trivial. To work directly with Lagrangians modulo full derivatives we shall replace $Cyc(U(I))$ by $Cyc(U(I))_{V}=H_0(\tilde YM,U(I))$.

There is an additional subtlety when we deform a Lagrangian $\L(\nabla,\xi)\rightarrow \L(\nabla,\xi)+\alpha' \L'(\nabla,\xi)$ by $\L'(\nabla,\xi)$. If $\L'(\nabla,\xi)=0$ on critical points of $\L(\nabla,\xi)$ then $\L(\nabla,\xi)+\alpha'\L'(\nabla,\xi)$ can be transformed to $\L(\nabla,\xi)$ by a field redefinition. In the algebraic language the space of on shell Lagrangians is equal to $H_0(\tilde YM,U(I)/(\tilde v_m,\tilde \chi_{\alpha}))=H_0(YM,U(TYM))=H_0(YM,\Sym(TYM))$, with $\tilde v_m,\tilde \chi_{\alpha}$-defining relations of $YM$ .


A simple way to deform $YM$ is to deform original $YM$-LLagrangian. One of the conclusions in \cite{M3} was that Yang-Mills equations  admit  deformations of which are not Euler-Lagrange . It is natural then to impose the condition {\bf Lg}:

Suppose we have a deformation $L_{\alpha'}\in \mathbf{L}(\susy({\alpha'}),TYM({\alpha'}))$. Then the induced deformation of $YM_{\alpha'}$ should have the relations $\frac{\delta {\cal L}_{\alpha'}}{\delta v_i}$, $\frac{\delta {\cal L}_{\alpha'}}{\delta \chi^{\alpha}}$ coming  from a Lagrangian ${\cal L}_{\alpha'}$.
\begin{remark}\label{R:dfpqmjh}
Abstract variational derivatives are uniquely defined by a property that
\begin{equation}
\begin{split}
&\Bigg(\frac{\delta \L'}{\delta v_i}\Bigg)(\nabla,\xi)=\frac{\delta \L'(\nabla)}{\delta \nabla_i}\\
&\Bigg(\frac{\delta \L'}{\delta \chi^{\alpha}}\Bigg)(\nabla,\xi)=\frac{\delta \L'(\nabla)}{\delta \xi^{\alpha}}
\end{split}
\end{equation}
\end{remark}

A similar definition can be given for the class of deformations $\mathbf{A}(\susy({\alpha'}),U(TYM)({\alpha'}))$. Indeed the algebra $U(YM)_{\alpha'}$ in this setup is a subalgebra of $U(L)_{\alpha'}$ generated by $\mu(\susy_2)$ and $U(TYM)_{\alpha'}$. The rest  parallels to  $\mathbf{L}(\susy({\alpha'}),TYM({\alpha'}))$.

Let us say a few words about possible homological interpretation of a statement that $\susy$ closes on shell in N=1 D=10 Yang-Mills theory. Indeed the space of functions (functionals) on the space of fields as we know can be substituted by $Cyc(U(I))$ or $\Sym(Cyc(U(I)))$ if we would like to work with multiple products of traces. The on shell functional are $H_0(TYM,U(TYM))$ ($\Sym(H_0(TYM,U(TYM)))$). The Lie algebra $L$ acts on $TYM$ by commutators. It continues on $H_0(TYM,U(TYM))$ and factors through $\susy=L/TYM$.

The same construction goes through for $\mathbf{A}$, $\mathbf{L}$ deformations and infinitesimal deformations. The key moment is that $TYM,U(TYM)$ are rigid. Thus the condition of infinitesimal on shell closure of $\susy$ is automatically satisfied for infinitesimal deformations of $\mathbf{A}$, $\mathbf{L}$ type.

\section{Deformation complexes $U(TYM)\otimes S$ and $TYM\otimes S$.}\label{S:hgdxsw}
In this section we start investigation of deformation cohomology  $H^k(L,TYM)$ and $H^k(L,U(TYM))$, introduced in the section (\ref{S:ibdhx}).

Section \ref{S:Lquad} offers some simplifications of deformation complexes. In the next few paragraphs we explain how results of section \ref{S:Lquad} can be adopted for our purposes.

\begin{definition}
Define a projective variety $\Q \subset \mathbf{P}^{15}$ by equations
\begin{equation}\label{E:equatdf}
r_i=\Gamma_{\alpha\beta}^i\lambda^{\alpha}\lambda^{\beta}=0
\end{equation}
\end{definition}
\begin{proposition}\label{P:dafystdf}
The algebra of homogeneous functions $S=\mathbb{C}[\lambda^1,\dots,\lambda^{16}]/(r_i)$ is Koszul (see \cite{Bezr} ). 
By \cite{MSch} and \cite{MSch2}  $U(L)=S^!$
\end{proposition}
\begin{proposition}\label{P:tqydxc}
The deformation cohomology $H^k(L,U(TYM))$ is equal to the $k$-th cohomology of the complexes $U(TYM)\otimes S$. The differential is a commutator with element 
\begin{equation}\label{E:uwhswtyrt}
e=\lambda^{\alpha}\theta_{\alpha}.
\end{equation} The cohomological grading coincides with the grading of $S$-factor. The total degree is preserved by $d$. The complex $U(TYM)\otimes S$ splits according to degree. A finer splitting can be achieved by identifying $U(TYM)=\bigoplus_{i\geq 0}\Sym^j(TYM)$.
\begin{equation}\label{E:fdgisev}
\begin{split}
&\Sym^j(TYM)_{n}\otimes S_0\rightarrow \Sym^j(TYM)_{n+1}\otimes S_1 \rightarrow\dots 
\end{split}
\end{equation}


\end{proposition}
\begin{proof}
By remark (\ref{R:wush}) the deformation cohomology $H^k(L,U(TYM))$ is equal to $H^k(U(L),U(TYM))$. The bimodule $U(TYM)$ has left $U(L)$- action induced by adjoint action of $L$. The right $U(L)$- action is induced by the trivial $L$-action. A similar $U(L)$-bimodule structure exists on $U(TYM)$. The propositions \ref{P:homolK}, (\ref{P:dafystdf}) applied to $U(L)$ and our modules  finish the proof.
\end{proof}
\begin{proposition}\label{P:iqwwst}
Homology $H_k(L,U(TYM))$ are equal to the cohomology of the complex $U(TYM))\otimes S^*$. The space $S^*=\bigoplus_{n\geq 0} S^*_n$ is a bimodule dual to  $S$. The differential is a commutator with element $e$ ((\ref{E:uwhswtyrt})). The homological degree coincides with the grading of  $S^*$-factor. The complex $U(TYM))\otimes S^*$ also splits :
\begin{equation}\label{E:homol}
\begin{split}
&\Sym^j(TYM)_{3j}\otimes S^*_{m}\rightarrow\dots \rightarrow  
\Sym^j(TYM)_{3j+m}\otimes S^*_{0}
\end{split}
\end{equation}

\end{proposition}
\begin{proof}
Is similar to the proof of proposition (\ref{P:tqydxc}).
\end{proof}

The complex group $Spin(10, \mathbb{C})$ acts transitively on $\Q $ ; 
the stable subgroup of a point is a parabolic subgroup $P$. To describe the
Lie algebra $\mathfrak{p}$ of $P $ we notice that the Lie algebra $
\mathfrak{so}(10, \mathbb{C}) $ of $SO(10, \mathbb{C}) $ can be identified
with ${\Lambda}^2(V) $ (with the space of antisymmetric tensors ${\rho}_{a
b} $ where $a, b=1, \dots, 10$). The vector representation $V$ of $SO(10, 
\mathbb{C}) $ restricted to the group $GL(5, \mathbb{C}) \subset SO(10, 
\mathbb{C}) $ is equivalent to the direct sum $W \oplus W^* $ of vector and
covector representations of $GL(5, \mathbb{C}) $. The Lie algebra of $SO(10, 
\mathbb{C}) $ as vector space can be decomposed as ${\Lambda}^2(W)+ 
\mathfrak{p} $ where $\mathfrak{p }=(W \otimes W^*) + \Lambda^2(W^*) $
is the Lie subalgebra of $\mathfrak{p}$. Using the language of generators we
can say that the Lie algebra $\mathfrak{so}(10, \mathbb{C}) $ is generated by
skew-symmetric tensors $m_{ab}, n^{ab} $ and by $k_a^b $ where $
a, b=1, \dots, 5 $. The subalgebra $\mathfrak{p}$ is generated by $k_a^b $
and $n^{ab} $. Corresponding commutation relations are 
\begin{align}
& [m, m^{\prime}]=[n, n^{\prime}]=0 \\
&[m, n]_a^b=m_{ac}n^{cb} \\
&[m, k]_{ab}=m_{ac}k_b^c+m_{cb}k_a^c \\
&[n, k]_{ab}=n^{ac}k_c^b+n^{cb}k_c^a
\end{align}


The complex  group $Spin(10)$ contains a two-sheet cover $\tilde{GL}(5)$ of $GL(5)$. Denote by $W$ the fundamental representation of $GL(5)$.  $\tilde GL(5)$ is the minimal cover  on which $det(g)^{\frac{1}{2}}, g\in \tilde GL_5$ becomes a single-valued representation. We denote it by $det(W)^{\frac{1}{2}}$. We denote $\tilde P=\tilde GL(5)\ltimes \Lambda^2(W^*)$ the two-sheet cover of $P$. The space $W \otimes det(W)^{-\frac{1}{2}}$ is $\tilde P$-representation.

Denote the bundle on $\Q $ induced from $W \otimes det(W)^{-\frac{1}{2}}$ by ${\cal W}$. The line bundle induced from $det(W)^{\frac{i}{2}}$ by $\O(i)$. A notation $\F\otimes \O(i)=\F(i)$  is common in algebraic geometry . 

%
The vector bundles $\Lambda^j{\cal W}(i)$ are $Spin(10)$ homogeneous. Thus the cohomology $H^k(\Q ,\Lambda^j{\cal W}(i))$ are $Spin(10)$-representations.

\begin{proposition}\label{P:azbush}
There is a long exact sequence connecting $H^k(L,U(TYM))$ and $H_k(L,U(TYM))$:
\begin{equation}\label{E:ywers}
\begin{split}
&\dots \rightarrow H_{3-i,a-8}(L,\Sym^j(TYM))\overset{\delta}{\rightarrow} H^{i,a}(L,\Sym^j(TYM))\rightarrow\\
& \rightarrow H^{i+a-3j}(\Q ,\Lambda^j(W)(3j-a))\overset{\iota}{\rightarrow}  H_{2-i,a-8}(L,\Sym^j(TYM))\rightarrow\dots
\end{split}
\end{equation}

\end{proposition}
\begin{proof}
See section (\ref{S:localiz}).
\end{proof}

\begin{proposition} Denote a sheaf of local holomorphic sections of $\Lambda^j{\cal W}(i)$ by the same symbol.
\begin{equation}\label{E:hdytsf}
\begin{split}
&i\geq 0\\
&H^0(\Q ,{\cal W}(i+1))=[1,0,0,0,i],\quad H^{10}(\Q ,{\cal W}(-8-i))=[0,0,0,i,1],\\
&H^0(\Q ,\Lambda^2{\cal W}(2+i))=[0,1,0,0,i],\quad H^{10}(\Q ,\Lambda^2{\cal W}(-8-i))=[0,0,1,i,0], \\
& H^9(\Q ,\Lambda^2{\cal W}(-6))=[0,0,0,0,0]\\
&H^0(\Q ,\Lambda^3{\cal W}(3+i))=[0,0,1,0,i],\quad H^{10}(\Q ,\Lambda^3{\cal W}(-7-i))=[0,1,0,i,0], \\
& H^1(\Q ,\Lambda^3{\cal W}(1))=[0,0,0,0,0]\\
&H^0(\Q ,\Lambda^4{\cal W}(3+i))=[0,0,0,1,i],\quad H^{10}(\Q ,\Lambda^4{\cal W}(-6-i))=[1,0,0,i,0], \\
&H^0(\Q ,\Lambda^5{\cal W}(3+i))=[0,0,0,0,i],\quad H^{10}(\Q ,\Lambda^5{\cal W}(-5-i))=[0,0,0,i,0], 
\end{split}
\end{equation}
denote ${\cal F}(t)=\sum_{i\geq 0}dim(H^0({\cal F}(i)))t^i$. Then 
\begin{equation}\label{E:hdytsf1}
\begin{split}
&\O(t)=\frac{1+5t+5t^2+t^3}{(1-t)^{11}}\\
&{\cal W}(t)=\frac{10+34t+16t^2}{(1-t)^{11}}\\
&\Lambda^2{\cal W}(t)=\frac{45+65t+11t^2-t^3}{(1-t)^{11}}\\
&\Lambda^3{\cal W}(t)=\frac{120-120t+330t^2-462t^3+462t^4-330t^5+165t^6-55t^7+11t^8-t^9}{(1-t)^{11}}\\
&\Lambda^4{\cal W}(t)=\frac{16+34t+10t^2}{(1-t)^{11}}
\end{split}
\end{equation}
%
\end{proposition}
\begin{proof}
The proof  reduces to a simple but tedious application of Borel-Weyl-Bott theorem, which was facilitated by a use of LiE program.
\end{proof}

The Lie algebra $TYM$  has a lowest graded component in the degree three (see (\ref{E:cxvfbw})). It gives a nontrivial subspace in $H_0(L,TYM)$. Similarly the subspaces $\Lambda^j(L_3)$ isomorphically map into $H_{0,3j}(L,\Sym^j(TYM))$ - subspace of degree $3j$.
\begin{proposition}\label{P:qideesa}
The following maps and inclusions are isomorphisms for $i\geq 4$.

\begin{equation}\label{E:jajqud}
\begin{split}
&H^{10}(\Q ,\Lambda^1({\cal W})(-8-i))=[0,0,0,i,1]\overset{\iota_1}{\rightarrow} H_{i,i+3}(L,TYM)\subset\\ &\subset  H_{i}(L,TYM)\\
&H^{10}(\Q ,\Lambda^2({\cal W})(-8-i))=[0,0,1,i,0]\overset{\iota_2}{\rightarrow} H_{i,i+6}(L,\Sym^2(TYM)) \subset\\ &\subset  H_{i}(L,\Sym^2(TYM))\\
&H^{10}(\Q ,\Lambda^3({\cal W})(-8-i))=[0,1,0,i+1,0]\overset{\iota_3}{\rightarrow} H_{i,i+9}(L,\Sym^3(TYM))\subset\\  &\subset H_{i}(L,\Sym^3(TYM))\\
&H^{10}(\Q ,\Lambda^4({\cal W})(-8-i))=[1,0,0,i+2,0]\overset{\iota_4}{\rightarrow} H_{i,i+12}(L,\Sym^4(TYM))\subset\\  &\subset H_{i}(L,\Sym^4(TYM))\\
&H^{10}(\Q ,\Lambda^5({\cal W})(-8-i))=[0,0,0,i+3,0]\overset{\iota_5}{\rightarrow} H_{i,i+15}(L,\Sym^5(TYM))\subset\\   &\subset H_{i}(L,\Sym^5(TYM))\\
&0=H_{i}(L,\Sym^j(TYM)),\quad j\geq 6
\end{split}
\end{equation}
\end{proposition}
\begin{proof}
Homology and cohomology of any Lie algebra are equal to zero in negative degree. $\Lambda^j({\cal W})=0$ for $j\geq 6$. These observations together with long exact sequence  (\ref{E:ywers}) are sufficient for the proof.
\end{proof}

\section{Classification of deformation cocycles in \\  $H^2(L,\Sym(TYM))$}\label{SS:dvausdu}
We make a classification of deformation cocycles in  $H^2(L,\Sym(TYM))$ by it relation kernel and image of maps $\delta$ (\ref{E:ywers}) and $d^L_{dR}$ \ref{E:DDDwwwDDD}.

\subsection{ Exceptional classes that are not in the image of $\delta$- type {\bf a} .}\label{S:gdujdsu}

These are the classes that span linear representations:
$[0,0,0,0,2]\subset H^{2,-2}(L,\mathbb{C})$, $[1,0,0,0,1]\subset H^{2,1}(L,TYM)$, $[0,1,0,0,0]\subset H^{2,4}(L,\Sym^2(TYM))$, which are preimages of (\ref{E:dvfsdhh}). There is also one sporadic (coming from higher cohomology): $c_{2,8}\in H^{2,8}(L,\Sym^3(TYM))$- a preimage of one of (\ref{E:jauuyd}).
\begin{proposition}
In the complex $S\otimes \Sym^3(TYM)$ the class $c_{2,8}$ is represented by an element $\Gamma_{\alpha\beta}^{i_1,\dots,i_5}\Gamma_{\gamma\delta i_1i_2i_3}\lambda^{\alpha}\lambda^{\beta}\otimes \chi^{\gamma}\bullet \chi^{\delta}\bullet F_{i_4i_5}\in S_2\otimes \Sym^3(TYM)_{10}$.
\end{proposition}
\begin{proof}
It is easy to see that $\Sym^3(TYM)_{10}=\Lambda^2(L_3)\otimes L_4$. The elements $c_{2,8}$ is the only $Spin(10)$-invariant element in $S_2\otimes \Lambda^2(L_3)\otimes L_4$.
\end{proof}
\begin{remark}
This  deformation cocycle was analyzed in \cite{Berg}, \cite{Cederwall} in connection with $\L'_{I}$.
\end{remark}

We leave as an exercise for the reader to check that all other classes belong to the image of $\delta$.(Hint: use long exact sequence (\ref{E:ywers}))

\subsection{Classes that are in the image of $\delta$.}\label{S:ddgwshcx}

These classes have their origin in homology group $H_1(L,\Sym^j(TYM))$. In the section (\ref{S:gqagbj}) we shall introduce an equivariant version of  Connes differential 
\begin{equation}\label{E:dfadsgfhc}
\dots \overset{d^L_{dR}}{\rightarrow}H_i(L,\Sym^j(TYM))\overset{d^L_{dR}}{\rightarrow} H_{i+1}(L,\Sym^{j-1}(TYM))\overset{d^L_{dR}}{\rightarrow}\dots
\end{equation}
 It satisfies $(d^L_{dR})^2=0$. According to proposition (\ref{C:gfdcbeyh}) $d^L_{dR}$ is zero on \\ $H_1(L,\Sym^j(TYM))$. It means that all elements of the later group are $d^L_{dR}$ cocycles and  can be classified with respect to  $d^L_{dR}$.

{\bf Nontrivial $d^L_{dR}$ cocycles in $H_1(L,\Sym^j(TYM))$-type b.}

This linear space according to corollary (\ref{C:gfddxfc}) is a direct sum $A^1_j+B^1_j$.
The content of $B^1_j$ is  written down in the lower part of (\ref{E:qjxhwyx}). The elements of these representations are mapped to nontrivial deformation classes in $H^2(L,\Sym(TYM))$ by the map $\delta$. 

The classes  $\gamma_{1,4}\in H_1(L,TYM), \gamma_{1,12}\in H_1(L,\Sym^3(TYM))$ are the only $Spin(10)$-invariants and are given by explicit formulas (\ref{E:dfdsfyh}).
Due to uniqueness of such classes and degrees counting we can identify the deformation $\delta\gamma_{1,4}$ with a deformation corresponding to infinitesimal Lagrangian $\L'_{II}$. The deformation $\delta\gamma_{1,12}$ corresponds to the Lagrangian $\L'_{III}$ which yet to be constructed (some information can be found in section (\ref{S:xaadbr})).
The classes $A_j^i$ according to  (\ref{C:gfddxfc}) are mapped by $\delta$ to zero, therefore do not concern us.

{\bf  $d^L_{dR}$ trivial cocycles in $H_1(L,\Sym^j(TYM))$- type c.}

These are elements in $H_1(L,\Sym^j(TYM))$ of the form $d^L_{dR}\gamma$, where $\gamma \in H_0(L,\Sym^{j+1}(TYM))$. We have an infinite-dimensional space of such elements.

\begin{remark}
Any element of $\Sym^{j+1}(TYM)$ produces (possibly zero ) element in $H_0(L,\Sym^{j+1}(TYM))$, for this we do not need to solve any equations, in contrast with elements of $H_1(L,\Sym^{j+1}(TYM))$.
\end{remark}
We defer discussion of $\mathbf{Lg}$ properties of constructed deformation cocycles until the end of section  (\ref{S:xaadbr}). Briefly we may say that all such deformations are of $\mathbf{Lg}$ type.

\section{A generating function for supersymmetric deformations}\label{S:gen}
The groups which govern deformations $H^{2}(L,U(TYM))$ or $H^{2}(L,TYM)$ have a grading. This enables us to form  generating functions:
\begin{equation}
\begin{split}
&a(t)=\sum_{i\geq -2}dim H^{2,i}(L,U(TYM))t^i\\
&l(t)=\sum_{i\geq 1}dim H^{2,i}(L,TYM)t^i
\end{split}
\end{equation}
In this section we procure  formulas for $a(t)$ and $l(t)$. The section will lack proofs: the reader can use \cite{M3} as guide to recover the missed points.

The main ingredients of our formulas are:
\begin{equation}
\begin{split}
&TYM(t)=\sum_{i\geq 3}dimTYM_it^i=\sum_{i\geq 3}\frac{(-1)^i}{i}\times\\ & \times \sum_{kd=i}\mu(d)(11+(-1)^{k-1}\{1+\frac{1}{(2+\sqrt 3)^k}+\frac{1}{(2-\sqrt 3)^k}\})t^i\\
&\overline{HC}_0(U(TYM))(t)=-\sum_{k\geq 1}ln(1-M((-1)^{k+1}t^k))\frac{\psi(k)}{k}\\
&U(TYM)(t)=\sum_{i\geq 3}dim U(TYM)_it^i=\frac{1}{1-M(t)}\\
&1-M(t)=\frac{1-4t+t^2}{(1-t)^4(1-t^2)^5}
\end{split}
\end{equation}
These have mostly a combinatorial origin.

$\mu(n)$ is the Mobius function. $\mu(n)=0$ if $n$ has  repeated prime factors, $\mu(1)=1$, $\mu(n)=(-1)^k$ if $n$ is a product of $k$ distinct primes.

$\psi(n)$ is the Euler function.  is defined as the number of positive integers $\leq n$ that are relatively prime to $n$.
\begin{equation}
\begin{split}
&p_0(t)=\frac{1-5t+5t^2-t^3}{(1+t)^{11}}\\
&p_1(t)=t^3\frac{16-34t+10t^2}{(1+t)^{11}}-t^4\\
&p_2(t)=t^6\frac{120+120t+330t^2+462t^3+462t^4+330t^5+165t^6+55t^7+11t^8+t^9}{(1+t)^{11}}-45t^8\\
&p_3(t)=-t^8\frac{45-65t+11t^2+t^3}{(1+t)^{11}}+45t^8+10t^{10}-t^{12}-144t^{11}\\
&p_4(t)=t^{10}\frac{10-34t+16t^2}{(1+t)^{11}}-10t^{10}+144t^{11}+16t^{13}-126t^{14}\\
&p_5(t)=-t^{12}\frac{1-5t+5t^2-t^3}{(1+t)^{11}}+t^{12}-16t^{13}+126t^{14}
\end{split}
\end{equation}
The last set of formulas is an appropriate modification of  \ref{E:hdytsf1}.

Finally
\begin{equation}\label{E:fvdvbhads}
\begin{split}
&a(t)=\Bigg(126t^{-2}+144t+45t^4+t^8\Bigg)+\Bigg(t^{12}+144t^{19}+t^{20}+126t^{22}\Bigg)+\\
&+\Bigg(-\frac{t^8(1-t)^{10}}{(1+t)^6}\overline{HC}_0(U(TYM))(t)-t^8(\sum_{i=1}^5ip_i(t))\Bigg)=a+b+c\\
&l(t)=\Bigg(144t\Bigg)+\Bigg(211t^8\Bigg)+\Bigg(t^8p_1(t)-t^8p_0(t)TYM(t)\Bigg)=a'+b'+c'
\end{split}
\end{equation}
In $a(t)$ the summand $a$ accounts for elements of type {\bf a} from section (\ref{S:gdujdsu}), term $b$ for type {\bf b} , $c$ for type {\bf c}. The same interpretation holds for $l(t)$.

Finally we tabulated dimensions of $Spin(10)$-invariants in \\ $H^{2,deg}(L,\Sym^p(TYM))$ in the following table.  We use LiE program in our computations.
 \begin{equation}\label{T:dkodd}
\mbox{
\scriptsize{
$\begin{array}{|c|c|c|c|c|c|c|c|c|c}
 &4&8&12&16&20&24&28&32& deg\\\hline
1& & &1& &1&3&18&172&\dots\\\hline
2& & & & & & &13&281&\dots\\\hline
3& &1& & &1&2&20 &267&\dots\\\hline
4& & & & & & &1 &68 &\dots\\\hline
5& & & & & & &1 &17 &\dots\\\hline  
6& & & & & & & & &\dots\\\hline
\Sym^p(TYM)&\dots&\dots&\dots&\dots&\dots&\dots&\dots &\dots &\dots
\end{array}$
}
}
\end{equation}

This table is a superposition of three tables according to types introduced in section \ref{SS:dvausdu}. The entries {\bf a,b}  tables were computed in \ref{SS:dvausdu}. The  table of type {\bf c}  is the biggest but the most regular. It was computed through  Euler characteristic of $H_{\bullet,k}(L,\Sym^i(TYM))$. The main ingredients in the computation are:  the complex (\ref{E:dfadsgfhc}), propositions (\ref{P:qideesa}, \ref{P;mxswess}),(\ref{P:xiuwbx}, \ref{E:vhbdc}). 
 It is possible in principle to write a generating function for the numbers from the above table in terms of characters of few basic $Spin(10)$ representations and Adams operations.
\section{Relation between ordinary and supersymmetric deformations}\label{S:xaadbr}

In this section we shall characterize a place of supersymmetric deformations among  ordinary deformations. Ordinary (nonsupersymmetric ) deformations of $YM$ are deformations of class $\mathbf{L}(V({\alpha'}),TYM({\alpha'}))$ , $\mathbf{A}(V({\alpha'}),U(TYM)({\alpha'}))$ or a $Spin(10)$-equivariant version, where $V$ is an even part of $\susy$. As in the case of supersymmetric analogs such deformations are parametrized by $H^2(YM,TYM)$, $H^2(YM,U(TYM))$ or  $Spin(10)$-invariants of one of the spaces.

Since $U(TYM)=\Sym(TYM)$ as $YM$-modules in our study we shall investigate relations between supersymmetric and ordinary cohomology with coefficients in $\Sym^j(TYM)$.

There is an operation of restriction of chains 
\begin{equation}\label{E:djfdshvba}
res:H^m(L,\Sym^k(TYM))\rightarrow H^m(YM,\Sym^k(TYM))
\end{equation} 
It is the map $res$ what we shall concentrate on in this section. We shall characterize its  kernel  and image  . In our study  $i$ is equal to $2$.

We use a spectral sequence of extension $YM\subset L$. The $E^{mn}_1$-term is equal to 
\begin{equation}\label{E:dhupksquy}
C^m(L/YM,H^n(YM,\Sym^k(TYM)))\Rightarrow H^{m+n}(L,\Sym^k(TYM))
\end{equation} 
The classes which contribute to $H^{2}(L,\Sym^k(TYM))$ are subquotients of 
\begin{equation}
\begin{split}
&C^0(L/YM,H^2(YM,\Sym^k(TYM)))\\
&C^1(L/YM,H^1(YM,\Sym^k(TYM)))\\
&C^2(L/YM,H^0(YM,\Sym^k(TYM)))
\end{split}
\end{equation} 
The classes of the last two linear spaces are mapped to zero under map $res$ because of nontrivial $L/YM$-dependence.  In proposition (\ref{P:ududufgqex}) we prove that $H^0(YM,\Sym^k(TYM))=0$ for $k\geq 1$. Thus the last term makes a  contribution to the kernel of $res$ for $k=0$, degree $i=-2$. The classes in $C^1(L/YM,H^1(YM,\Sym^k(TYM)))$ by proposition (\ref{P:ududufgqex}) have degree $i=-3,-4$ for $k=0$ and   $i=0,1$ for $k=1$. For $k\geq 2$ the corresponding linear space is zero. From long exact sequence (\ref{E:ywers}) nontrivial classes in $H^{2,i}(L,\Sym^k(TYM))$ for mentioned above degrees and values of $k$ exist only for $k=0$, $i=-2$, $k=1$, $i=1$. The group $H^{2,-2}(L,\mathbb{C})$ is isomorphic to $[0,0,0,0,2]$,  $H^{2,1}(L,TYM)$ is isomorphic to $[1,0,0,0,1]$ as $Spin(10)$-representation.

\begin{proposition}
The kernel of map (\ref{E:djfdshvba}) ($m=2$) is nontrivial only if $i=-2$ $k=0$ and is equal to $[0,0,0,0,2]$; in degree $i=1$ $k=1$  and is equal to $[1,0,0,0,1]$
\end{proposition}
The deformations which belong to $\Ker(res)$ of odd degree are unphysical.
 We can interpret $\Ker(res)$ as deformation of action of supersymmetries on $YM$, while the algebra $YM$  is kept undeformed. It is easy to see that a deformation of degree $-2$ though changes $L$ does not affect the adjoint action of $\theta_{\alpha}(\alpha')$ on $YM(\alpha')$. Thus this deformation can also be discarded.
\begin{proposition}
The image of $res$ in $H^2(YM,\Sym^k(TYM))$ in $deg\geq 1$ can be characterized  as classes invariant with respect to $\susy$. 
\end{proposition}
\begin{proof}
We shall analyze higher differentials in this spectral sequence. There is  only one  possible  nontrivial differential acting on \\  $H^0(L/YM,H^2(YM,\Sym^k(TYM)))$ 
\begin{equation}\label{E:vhdvbhv}
d_2:H^0(L/YM,H^2(YM,\Sym^k(TYM)))\rightarrow H^2(L/YM,H^1(YM,\Sym^k(TYM)))
\end{equation}
An interpretation of this map is the following. An element \\ $\gamma\in H^2(YM,\Sym^k(TYM))$ defines infinitesimal deformation of $YM$. If it belongs to \\ $H^0(L/YM,H^2(YM,\Sym^k(TYM)))\subset H^2(YM,\Sym^k(TYM))$ then each supersymmetry can be modified to $\theta_{\alpha}+\alpha'\theta'_{\alpha}$ such that it defines a derivation of the infinitesimal deformation. A commutator of such two derivations defines an even derivation of $YM$, whose $\alpha'$ coefficient $\tau_{\alpha\beta}$ for fixed $\alpha\beta$ is an element of $H^1(YM,\Sym^k(TYM))$ (recall a discussion in section (\ref{S:ibdhx}) about derivations). The later group was computed in (\ref{E:vhdvbhv}). It is nontrivial for $k=0,1$ and for $k=1$ can be identified with $\susy$. The whole object $\tau_{\alpha\beta}$ is an element of $H^2(L/YM,H^1(YM,\Sym^k(TYM)))$. The condition that $\gamma\in \Ker d_2$ is equivalent to a property that infinitesimally the action of an even translation receives a  correction which can be eliminated  by a field redefinition. It does not mean that we can do it simultaneously for all translations.


In the table below the reader may see degrees $i$ of nonzero components of source and target (\ref{E:vhdvbhv}) for different values of $k$.
\begin{center}\label{E:yufwd}
\scriptsize{
$\begin{array}{ccc}
k&H^0(L/YM,H^2(YM,\Sym^k(TYM)))&H^2(L/YM,H^1(YM,\Sym^k(TYM)))\\\hline
0&-6,-5\ \ \ \ \ \ \   &-5,-4 \\
1&-2,-1,0,\dots&-1,0\ \    \\
2&1,2,\dots\ \ \    & \\
3&4,5,\dots\ \ \    & \\
\dots&\dots\ \ \ \ \ \ \  &
\end{array}$
}
\end{center}
Due to homogeneity with respect to $deg$ the map could be  nontrivial only for $k=0$ in degree $-5$, for  $k=1$ in degree $0$. In both cases we do not have $Spin(10)$-invariant deformations.
\end{proof}

We need to remind to the reader(trivial from the point of view of the calculus of variations) relation between Lagrangians and deformations.

The first fact is that the Lie algebra $YM$ is an algebra with Poincare duality (see \cite{MSch2}). It means that for any graded module $N$ we have 
\begin{equation}\label{E:poincare}
H^{i,j}(YM,N)=H_{3-i,j+8}(YM,N),
\end{equation} the second index is the degree. It allows us to identify
\begin{equation}
H^{2}(YM,U(TYM))\overset{P}{\rightarrow}H_{1}(YM,U(TYM))=H_{1}(YM,\Sym(TYM))
\end{equation}

The second fact is that there is a Connes differential defined in (\ref{S:gqagbj}).
\begin{equation}
\dots\rightarrow H_i(YM,\Sym^k(TYM))\overset{d_{dR}^{YM}}{\rightarrow} H_{i+1}(YM,\Sym^{k-1}(TYM))\rightarrow \dots 
\end{equation}
 It allows to map
\begin{equation}\label{E:dvdsbn}
 H_{0,i}(YM,\Sym^k(TYM))\xrightarrow{var=P^{-1}\circ d_{dR}^{YM}} H^{2,i-8}(YM,\Sym^{k-1}(TYM)) 
\end{equation}

It is easy to see that $var$ has a simple variational interpretation. Yang-Mills theory has a Lagrangian $\L(\nabla,\xi)$. Pick an element $\L'\in H_0(YM,U(TYM))$ and define a deformed Lagrangian $\L(\nabla,\xi)+\alpha'\L'(\nabla,\xi)$. The deformed $YM$ algebra according to  remark (\ref{R:dfpqmjh}) has relations $\frac{\delta \L+\alpha'\L'}{\delta v_i}$, $\frac{\delta \L+\alpha'\L'}{\delta \chi^{\alpha}}$. It is plausible and easy to check that cocycle of such deformation is $var(\L')$.

This enables us to translate into algebraic language condition $\mathbf{Lg}$. 

Suppose we have a deformation with a cocycle  $c\in H^2(L,U(TYM)$. Then $c$ is $\mathbf{Lg}$-type if $res(c)$ belongs to the image of $var$ and $var^{-1}\circ res(c)$ is the infinitesimal Lagrangian.

Remark (\ref{R:fdsdsaq}) asserts that any cocycle in $H^2(L,U(TYM)^{Spin(10)}$ is of type $\mathbf{Lg}$ in the above sense.

Now we know everything to explain formulas (\ref{E:osdodsda}) and (\ref{E:knhbjh}). The factor $t^8$ corresponds to degree of Poincare duality map (\ref{E:poincare}), because the later is involved in the definition of $var$. The terms which are subtracted from $a(t)$ define a generating function of $\Ker res$. The relation of $l(t)$ and $\tilde{l}(t)$ is essentially the same.

We can explain relation of tables \ref{T:dkodd1} and (\ref{T:dkodd}) along the same lines. The reader can see that the tables have the same entries, but a different numeration of rows and columns. The tables have the same entries because as we already know $res$ is injective on $H^2(L,\Sym(TYM))^{Spin(10)}$ and all $Spin(10)$-invariants are in the image of $var$. The change of row numbering is due to increase $k\rightarrow k-1$ in the definition of $var$ in (\ref{E:dvdsbn}).  The change in column numbering comes from two sources. Firstly, the map $var$ changes degree $deg$ by $8$. Secondly in transition $deg\rightarrow deg_{\alpha}$ we use a formula (\ref{E:aaghhs}).

In the introduction we made a claim that Lagrangian $\L_{IV}$ can be constructed through the operator $\epsilon^{\alpha_1\dots\alpha_{16}}\theta_{\alpha_1}\dots \theta_{\alpha_{16}}$. Let us elaborate on this claim. From our discussion at the end of section (\ref{S:ibdhx}) about Lagrangians it should be clear that the right algebraization of $trn(\nabla,\xi)$ is an element of $Cyc(U(I))$. The last group maps to $H_0(L,U(TYM))$.

One of the ways to produce supersymmetry invariant infinitesimal Lagrangians   would be to take an element $\L' \in H_0(L,\Sym^k(TYM))$, map it to \\ $H_1(L,\Sym^{k-1}(TYM))$ via  differential $d^{L}_{dR}$. Then use map $\delta$ to transport it  to $H^2(L,\Sym(TYM))$. Then by $res$ to $H^2(YM,\Sym(TYM))$. After that take a preimage  $var$ and end up in desired group $H_0(YM,\Sym^k(TYM))$

One can do differently and use proposition (\ref{P:hygdscv}). We know that $res\circ \delta \circ d^{L}_{dR}\L'=P^{-1}\circ d^{YM}_{dR}\circ P\circ res\circ\delta \L'$.
This enables us to claim that  $P\circ res\circ\delta \L'$  is an infinitesimal deformation of the Lagrangian corresponding to cocycle $\delta \circ d^{L}_{dR}\L'$.
Due to  proposition (\ref{P:hygdscv}) we may claim that $P\circ res\circ\delta \L'=\epsilon^{\alpha_1\dots\alpha_{16}}\theta_{\alpha_1}\dots \theta_{\alpha_{16}}\L'$

%
%
%
%

\section{Appendix}

\subsection{Cohomology of quadratic algebras}\label{S:Lquad}
Material of this section is necessary for justification of reduction of big deformation complex $C(L,\Sym(TYM))$ to a much smaller complex $S\otimes \Sym(TYM)$. We use this in section (\ref{S:hgdxsw})

 It is well known that tangent space to the space of $A_{\infty}$ deformations of an algebra $A$ is governed by Hochschild cohomology $HH^{\bullet}(A,A)$. We define an object slightly more general - Hochschild cohomology with coefficients in a bimodule $M$ . We denote it  by $HH^{\bullet}(A,M)$. It is a cohomology of a complex 
\begin{equation}\label{E:WWD}
C^n(A,M)=Hom(A^{\otimes n},M) \mbox{ with differential} d:C^n(A,M)\rightarrow C^{n+1}(A,M)
\end{equation} defined by the formula 
\begin{align}\label{E:lovnms}
&d(c)(a_0, \dots ,a_n)=a_0c(a_1, \dots ,a_n)+\\ \notag
&+\sum_{i=1}^n(-1)^ic(c_0, \dots ,a_{i-1}a_i, \dots , a_n)+(-1)^{n+1}c(a_0, \dots .a_{n-1})a_n
\end{align}

If the algebra $A$ and the module $M$  are  graded ($A=\bigoplus_{i=0}^{\infty} A_i$ with $A_0=\mathbb{C}$, $M=\bigoplus_{i\in \mathbb{Z}} M_i$, then the complex $C^i(A,M)$ and cohomology  are graded $C^{i,k}=Hom^k(A^{\otimes i},M)$ by degree of a map . In the  superscript  $HH^{i,j}(A, M)$ the first is cohomological index, the second is the degree.

\begin{remark}\label{R:wush}
This construction is related to the Lie algebra cohomology, described in section (\ref{S:ibdhx}). Indeed if $A=U(\mathfrak{g})$, a bimodule  $M$ defines an adjoint $\mathfrak{g}$-module, which we denote by $M^{ad}$. The action of $l\in \mathfrak{g}$, $m\in M^{ad}$, then $lm=lm-ml$.

According to \cite{McL} there is an isomorphism $HH^k(U(\mathfrak{g}),M)=H^k(\mathfrak{g},M^{ad})$.
\end{remark}

We also shall use a dual construction- homology groups $HH_i(A,M)$. Here it is brief description:
Denote $C_n(A,M)=A^{\otimes n}\otimes M$ the groups which constitute a complex $d:C_n(A,M)\rightarrow C_{n-1}(A,M)$ with a differential :
\begin{equation}\label{E:ucyegx}
\begin{split}
&d(m\otimes a_1 \otimes\dots \otimes a_n)=ma_1\otimes a_1\otimes\dots \otimes a_n +\\
&+\sum_{i=1}^{n}(-1)^im\otimes a_1 \otimes\dots\otimes a_ia_{i+1}\otimes\dots\otimes a_n\\
&+(-1)^na_nm\otimes a_1 \otimes\dots \otimes a_{n-1}
\end{split}
\end{equation}

In our brief exposition of quadratic algebras we closely follow \cite{Qalg}.
\begin{definition}
A graded algebra $A=\bigoplus_{i\geq 0} A_{i}$ is
called a quadratic  if $A_{0}=\mathbb{C}, $  $W=A_{1}$ 
generates $A$ and all relations follow from quadratic relations $\sum\limits_{i, j}r_{ij}^{k}\lambda^{i}\lambda^{j}=0$ where $\lambda^{1}, \dots , \lambda^{\dim W}
$ is a basis of $W$ The space of quadratic relations (the
subspace of $W\otimes W$ spanned by $r_{ij}^{1}, r_{ij}^{2}, \ldots $)
 will be denoted by $R.$
\end{definition}

 We  see that $A_{2}=W\otimes W/R$ and 
$A$ is a quotient of free algebra (tensor algebra) $T(W)$ with
respect to the ideal generated by $R.$ 
\begin{definition}
The dual quadratic algebra $A^{!}$ is defined as a quotient of a free algebra $T(W^*)$
 generated by $W^{\ast }$  (with a basis $\theta_{1},\dots, \theta_{\dim W}$ dual to $\lambda^{1},\dots,\lambda^{\dim W}$) by the  ideal
generated by $R^{\perp }\subset W^{\ast }\otimes W^{\ast }$ (here $%
R^{\perp }$ stands for the subspace of $W^{\ast }\otimes W^{\ast
}=\left(W\otimes W\right) ^{\ast }$ that is orthogonal to $R\subset
W\otimes W$). 
\end{definition}

There is an element $e$ in the tensor product $A_1\otimes A^!_1=W\otimes W^*=End(W)$. It corresponds to  the identity element in $End(W)$. Equation $e^2=0$ is a direct corollary of orthogonality relations between $R$ and $R^{\perp}$. The space $A^{!*}$ is dual bimodule. The  left multiplication on $e$ defines    an operator 
\begin{equation}
\dots \overset{d}\rightarrow A_i\otimes A^{!*}_{n-i} \overset{d}\rightarrow A_{i+1}\otimes A^{!*}_{n-i-1}  \overset{d}\rightarrow \dots
\end{equation} 
It automatically satisfies $d^2=0$. The complex $A\otimes A^{!*}$ is called Koszul complex of $A$ (the cohomological degree coincides with $A^{!*}$ grading)

\begin{definition}\label{P:prk}

A quadratic algebra $A$ is called Koszul if $K_n(A)$ is acyclic for $n>0$.
\end{definition} 

For Koszul  algebra $A$ there is  a more economical complex then \ref{E:WWD} suitable for computations of cohomology $HH^i(A,M)$. 
\begin{proposition}\label{P:homolK}
Suppose $A$ is a Koszul algebra and $M$ is $A$-module.
The cohomology of the complex $M\otimes A^!$ with a differential $d(a)=[e,a]$ is isomorphic to $HH^{i}(A,M)$. There is an appropriate refinement of this statement when $M$ is graded.
\end{proposition} 
\begin{proof}

 A minor modification of $A\otimes A^{!*}$ is the complex $A\otimes A\otimes A^{!*}$. The differential is of a  bicomplex $(A\otimes A\otimes A^{!*}, d_1, d_2)$. Let $e=\sum_{\alpha}\theta_{\alpha}\otimes \lambda^{\alpha}$. Denote $e_1=\sum_{\alpha}\theta_{\alpha}\otimes 1 \otimes\lambda^{\alpha}$ and $e_2=\sum_{\alpha}1 \otimes \theta_{\alpha} \otimes\lambda^{\alpha}$, then $d_1$ is a left multiplication on $e_1$ and $d_2$ is a right multiplication on $e_2$. Define a cohomological grading $A\otimes A\otimes A^{!*}$ by the degree of $A^{!*}$ factor.

\begin{lemma}
Suppose $A$ is a Koszul algebra.
The complex $A\otimes A\otimes A^{!*}$ is acyclic in all degrees, but zero. Zero cohomology is  equal to $A$.
\end{lemma}
\begin{proof}
The proof is based on studying spectral sequence associated with a bicomplex $(A\otimes A\otimes A^{!*}, d_1, d_2)$, which collapses due to Proposition (\ref{P:prk}).
\end{proof}

 Recall that there is an interpretation of groups $HH^{\bullet}(A,M)$ through resolutions.
\begin{definition}
Denote $A^{op}$ an algebra with linear space of $A$ and multiplication $a\times b=ba$.
\end{definition}
Any $A$-bimodule $M$ becomes left $A\otimes A^{op}$-module by the formula $(a\otimes b)m=amb$
\begin{lemma}\label{P:sdf} { \cite{CE}}
For any algebra $A$ and a bimodule $M$ there is an isomorphism of groups $HH^{i}(A,M)=Ext_{A\otimes A^{op}}^i(A,M)$. In the last group $A$ is understood as $A \otimes A^{op}$-module.
\end{lemma}

This lemma may have the following interpretation: take any projective resolution of $A \otimes A^{op}$-module $A$:
\begin{equation}\label{E:reso}
A\leftarrow P_0 \leftarrow P_1 \leftarrow \dots \leftarrow P_n \leftarrow\dots
\end{equation}
Then the cohomology of the complex $Hom_{A \otimes A^{op}}(P_n,M)$ is equal to $HH^{n}(A,M)$ (the reader may consult \cite{CE} or \cite{McL} for the relevant definitions and details).

We  can use a complex $A\otimes A\otimes A^{!*}$ as a resolution of $A$, where $P_n=A\otimes A\otimes A^{!*}_n$. It is a straightforward check that $Hom_{A \otimes A^{op}}(P_n,M)=A^{!*}_n\otimes M$. The differential $d(a)$ in this complex is a commutator of  $a$  with the canonical element  $e$. If $a \otimes b \in M\otimes A^!$, then the homogeneity degree $l=deg(a)-deg(b)$ is preserved by the differential $d$. Then  $$M\otimes A^!=\bigoplus_l (M\otimes A^!)_l$$ -is direct sum of smaller complexes spanned by elements of homogeneity degree $l$ which we denote by $(M\otimes A^!)_l$. 

\end{proof}

There is a version of this theory in homological setup.

\begin{proposition}
Suppose $A$ is a Koszul algebra and $M$ is $A$-bimodule. Then homology groups $H_i(A,M)$ is isomorphic to cohomology of the complex $M\otimes  A^{!*}_n $ with differential $d(a)=[e,a]$.
\end{proposition}
\begin{proof}
Similar to the proof of proposition \ref{P:homolK}. Use flat $A \otimes A^{op}$-resolution $A\otimes A\otimes A^{!*}$ of $A$. 
%
%
%
\end{proof}

\subsection{Localization of $\Sym^j(TYM)\otimes S$ and $\Sym^j(TYM)\otimes S^*$}\label{S:localiz}
In this section we prove proposition (\ref{P:azbush}).

.

\begin{proposition}\label{P:siwjfue}\cite{MSch2}
The cohomology  $H^i(\Q ,{\cal O}(n))$ is equal to $S_n$, if $i=0$ and to $S_k^*$, if $n=-8-k,i=10, k\geq 0$. All other cohomology vanish
\end{proposition}

Let $N=\bigoplus_{i\geq 0}N_i$ be a graded $L$-module. Consider a complex of vector bundles

\begin{equation}\label{E:fdgisevdfd}
N_{0}(l)\rightarrow  N_{1}(l+1) \rightarrow\dots \rightarrow N_{k}(l+k)\rightarrow \dots 
\end{equation}
with differential defined by multiplication on element $e$ from (\ref{E:uwhswtyrt}). 
In (\ref{E:fdgisevdfd}) $N_n$ is understood as trivial vector bundle with a fiber $N_n$.

The (${\cal O}(l)$-twisted ) localization of (\ref{E:fdgisev}) or (\ref{E:homol}) is the complex (\ref{E:fdgisevdfd}) where $N_n=\Sym^j(TYM)_n$.
In such case by proposition (\ref{P:siwjfue}) the complex of global sections of (\ref{E:fdgisevdfd}) for appropriate $l$ and  a shift is identical to  (\ref{E:fdgisev}), whereas the complex of tenth cohomology of (\ref{E:fdgisevdfd}) for suitable $l$ and shift is equal to (\ref{E:homol}).

\begin{definition}\label{D:fgysvbc}
Denote the complex of vector bundles (\ref{E:fdgisevdfd}) by $\N(l)$. The bicomplex of Dolbeault of $(0,p)$- forms with coefficients in $\N(l)$ we denote by $\Omega\N(l)$. It has two differentials: the original $d$ and $\dbar$. The grading of the element in $\Omega^{0,p}N_{i}(l+i)$ is equal to $p+i$. The total complex we denote by $\Omega\N^{\bullet}(l)$.
\end{definition}
There are two mappings- an embedding 
\begin{equation}
i:H^0(\Q ,\N(l))^{\bullet}\rightarrow \Omega\N^{\bullet}(l)
\end{equation}
 and projection 
\begin{equation}
p:\Omega\N^{\bullet}(l)\rightarrow H^{10}(\Q ,\N(l))^{\bullet}[-10]
\end{equation}
\begin{proposition}
The map $p:\Omega\N^{\bullet}(l)/\Im^{\bullet}(i)\rightarrow H^{10}(\Q ,\N(l))^{\bullet}[-10]$ is a quasiisomorphism.
\end{proposition}
\begin{proof}
Consider a spectral sequence of a bicomplex for $\Omega\N^{\bullet}(l)/\Im^{\bullet}(i)$, in which we compute the cohomology of $\dbar$ first. 
The $E_1$-term has nonzero entries equal to $E_1^{10,i}=H^{10}(\Q ,{\cal O}(l+i))\otimes N_i$, whence the proof.
%
%
\end{proof}

\begin{corollary}\label{C:osadfx}
There is a long exact sequence of cohomology
\begin{equation}\label{E:delta}
\begin{split}
& \dots \rightarrow H^i(H^0(\Q ,\N(l)))\rightarrow H^i(\Q ,\Omega\N(l))\rightarrow\\
& \rightarrow H^{i-10}(H^{10}(\Q ,\N(l)))\overset{\delta}{\rightarrow} H^{i+1}(H^0(\Q ,\N(l)))\rightarrow \dots
\end{split}
\end{equation}
\end{corollary}

From now on we set $N$ to be equal to $\Sym^j(TYM)$.
Our next goal is to compute the cohomology of a fiber of $\Sym^j(\ctym)^{\bullet}(l))$ over a point $x\in \Q$. Due to $Spin(10)$ homogeneity, fibers over different points are isomorphic.  This will enable us to compute $H^i(\Q ,\Omega\Sym^j(\ctym)(l))$ completely. To do this we need to remind some facts about the manifold $\Q $.

It is good deal is known about $\Q $ (see \cite{MSch} for list of properties and references.) The main properties of $\Q $ are :

1. The complex dimension of $\Q $ is equal to 10.

2. As a real homogeneous manifold $\Q $ is equal to $SO(10,\mathbb{R})/U(5)$.

It follows from this that $\Q $ is smooth. The affine cone, defined by equation (\ref{E:equatdf}), is also smooth  away from the apex.


If we choose  $\lambda_0^{\alpha}$- a solution to (\ref{E:equatdf}), then $\theta=\sum_{\alpha}\lambda_0^{\alpha}\theta_{\alpha}$ satisfies $[\theta,\theta]=2\theta^2=0$. It can be used to define a differential $d$ on $L$ and on $U(L)$ by the formula $d(a)=[\theta,a]$. A vector $\lambda_0^{\alpha}$ is coordinates of  point $x$ on the cone $C\Q $.

One can define a one-dimensional $S$-bimodule $\mathbb{C}_x$ by specialization at $x\in C\Q $ with coordinates $\lambda_0^{\alpha}$. To emphasize $x$-dependence of the differential $d$ we denote it by $d_x$.

\begin{proposition}
$HH^i(S,\mathbb{C}_x)=H^i(U(L),d_x)$
\end{proposition}
\begin{proof}
This is a direct application of proposition \ref{P:homolK}, where $M=\mathbb{C}_x$.
\end{proof}

\begin{proposition}
Suppose $A$ is a ring of algebraic functions on affine algebraic variety. $\mathbb{C}_x$ is a one-dimensional bimodule, corresponding to a smooth point $x$. Then $HH^i(A,\mathbb{C}_x)=\Lambda^i(T_x)$, where $T_x$ is the tangent space at $x$.
\end{proposition}
\begin{proof}
This is a weak form of Hochschild-Kostant-Rosenberg theorem.
\end{proof}
\begin{corollary}\label{C:hgfjs}
$HH^i(S,\mathbb{C}_x)=H^i(U(L),d_x)=\Lambda^i(T_x)$, where $T_x$ is the tangent space to $C\Q $ at the point $x$, $x\neq 0$.
\end{corollary}
\begin{remark}\label{R:suwhd}
This can be interpreted in elementary terms. Suppose we would like to deform $d_x$ by deforming $x$. The condition 
\begin{equation}\label{E:equatdf1}
\Gamma_{\alpha\beta}^i\lambda_0^{\alpha}\xi^{\beta}=0 \quad i=1,\dots,10
\end{equation}
is the condition that a vector with coordinates $\xi^{\alpha}$ is tangent to $C\Q $ at $x$. This is precisely the condition that $[e_x,g]=0$, where $g=\sum_{\alpha}\xi^{\alpha}\theta_{\alpha}$. Multiplicative structure in $U(L)$ allows to multiply different $g$'s, spanning the exterior algebra $\Lambda^i(T_x)$. A less trivial fact established in (\ref{C:hgfjs}) is that this is how we can generate all the cohomology.
\end{remark}
\begin{corollary}\label{C:jsfdnsd}
$H^i(\Sym^i(L),d_x)=\Lambda^i(T_x)$ and all other cohomology equal to zero.
\end{corollary}
\begin{proof}
By Poincare-Birkhoff-Witt theorem $(U(L),d_x)=(\Sym(L),d_x)$. It means that we have a direct sum decomposition of subcomplexes $(\Sym(L),d_x)=\bigoplus_{n\geq 0}(\Sym^n(L),d_x)$.
In the previous paragraph we established that $T_x\subset H^1(L,d_x)$. By construction $\Lambda^i(T_x)\subset H^i(\Sym^i(L),d_x)$. By (\ref{C:hgfjs}) $\Lambda(T_x)$ exhaust all the cohomology.
\end{proof}

Our next goal is the computation of $H(\Sym^{j}(TYM),d_x)$. 

For this  is useful to know $Spin(10)$-representation content of the Lie algebra $L$. 

A general algorithm to determine isotopic component in $L$  is the following:

1. We know the representation content of homogeneous components of the algebra $S=\bigoplus_{n\geq0}S_n$
The spinor representation $S_1$ has the highest weight $[0,0,0,0,1]$. The $n$-th component $S_n$ (by Borel-Weyl-Bott theorem) is an irreducible representation of weight $[0,0,0,0,n]$. Koszul property of $S$ tells us that there is a series of acyclic complexes (differential is a left multiplication on $e$.)
\begin{equation}
U(L)_n\otimes S^*_0\leftarrow U(L)_{n-1}\otimes S^*_1 \leftarrow\dots \leftarrow U(L)_{0}\otimes S^*_{n}
\end{equation}
They can be used to compute $Spin(10)$-representation in  $U(L)_n$ inductively.

We can assume now that we know the content of all $U(L)_n$. The main formula which enables to recover the content of $L_n$ from $U(L)_n$ is the Poincare-Birkhoff-Witt isomorphism $U(L)=\Sym(L)$. It can be used to compute $L_n$ by induction.  Such computations are greatly facilitated by the use of LiE program. The low graded components of $L$ are

\begin{equation}
\begin{split}
& L_1=[0,0,0,1,0]\\
& L_2=[1,0,0,0,0]\\
& L_3=[0,0,0,0,1]\\
& L_4=[0,1,0,0,0]\\
&\dots
\end{split}
\end{equation}

\begin{proposition}
There are the following identifications of $\widetilde{GL}_5$-representations :
\begin{equation}\label{E:decomp}
\begin{split}
&L_1=\left(\mathbb{C}+\Lambda^2(W)+\Lambda^4(W)\right)\otimes det(W)^{-\frac{1}{2}}\\
&L_2=W+W^*\\
&L_3=\left(\mathbb{C}+\Lambda^2(W^*)+\Lambda^4(W^*)\right)\otimes det(W)^{\frac{1}{2}}=\\ &=\left(\mathbb{C}+W\otimes det(W)^{-1}+\Lambda^3(W)^{-1}\right)\otimes det(W)^{\frac{1}{2}}\\
&L_4=\Lambda^2(W)+\Lambda^2(W^*)+W\otimes W^*\\
&\dots
\end{split}
\end{equation}
\end{proposition}
\begin{proof}
Use a realization of spinor representation in even (odd) exterior powers of $W$ (Fock representation) described in \cite{Cartan}.
\end{proof}
\begin{proposition}
$H^3(TYM,d_x)=W\otimes det(W)^{-\frac{1}{2}}$, all other cohomology vanish
\end{proposition}
\begin{proof}

It not hard to describe the action of the differential $d_x$ on $L_n$ for small $n$, where the point $x$ is  invariant with respect to $SL_5$.
We  describe the differential $d_x$ explicitly using decomposition (\ref{E:decomp}). We included  only  those isotopic components on which $d_x$ is not zero. On such components the differential is uniquely defined up to a scalar factor
\begin{equation}
\begin{split}
&L_1\supset \Lambda^2(W)\otimes det(W)^{\frac{1}{2}}=W^*\otimes det(W)^{\frac{1}{2}} \overset{d_x}\rightarrow W^*\subset L_2\\
&L_2\supset W \overset{d_x}\rightarrow  W\otimes det(W)^{-\frac{1}{2}} \subset L_3\\
&L_3\supset det(W)^{\frac{1}{2}}+\Lambda^3(W)\otimes det(W)^{-\frac{1}{2}}\rightarrow \mathbb{C}+\Lambda^2(W^*)\subset W\otimes W^*+\Lambda^2(W^*)
\end{split}
\end{equation}

The complex $L_1/T_x\rightarrow L_2\rightarrow \dots$ is acyclic. If we truncate $L_1/T_x$ and $ L_2$ terms, the resulting complex will have cohomology equal to $d_x(L_2)=W\otimes det(W)^{-\frac{1}{2}}$.
\end{proof}
\begin{corollary}
The complex $(\Sym^{j}(TYM),d_x)$ has cohomology in degree $3i$  equal to $\Lambda^i(W\otimes det(W)^{-\frac{1}{2}})$.
\end{corollary}
\begin{proof}
Similar to the proof of corollary (\ref{C:jsfdnsd}).
\end{proof}

To find  cohomology of $(\Sym^{j}(TYM)\otimes S,d)$ we plan study linear spaces $\Lambda^j(W_x\otimes det(W_x)^{-\frac{1}{2}})$ as a family $x\in C\Q \backslash  \{0\}$. 

There is a natural identification of linear spaces $\Lambda^j(W_x\otimes det(W_x)^{-\frac{1}{2}})$ and $\Lambda^j(W_{\lambda x}\otimes det(W_{\lambda x})^{-\frac{1}{2}})$ $\lambda \in \mathbb{C}^{\times}$.
From this we conclude that the family $W_x\otimes det(W_x)^{-\frac{1}{2}}$ can be pushed to a vector bundle ${\cal W}$  on  $\Q $.
\begin{proposition}\label{P:qydcc}
$H^i(\Omega\Sym^j\ctym(l))=H^{i-3j}(\Q ,\Lambda^j{\cal W}(l))$
\end{proposition}
\begin{proof}
In corollary (\ref{C:jsfdnsd}) we found  that local (at a point) cohomology of \\ $\Sym^j\ctym(l)$ is equal to $\Lambda^j({\cal W})(l)[-3j]$. In fact the later vector bundle is a subbundle of $\Sym^j\ctym(l)$. Thus the embedding $\Omega \Lambda^j({\cal W})(l) \rightarrow \Omega\Sym^j\ctym(l)$ is a quasiisomorphism. 
\end{proof}

\begin{proof}{\bf of Proposition (\ref{E:ywers})}
One simply has to make necessary shift in cohomological degree and grading when identifies $H_{i,k}(L,\Sym^j(TYM))$ and $H^{\bullet}(H^{10}(\Q ,\Sym^j(TYM)(l)))$. The same applies to cohomology. Then use proposition (\ref{P:qydcc}) and (\ref{C:osadfx}).
\end{proof}

\begin{proof}{\bf of lemma (\ref{L:wsdsfsc})}
A plan is to use multiplicative action of $H^i(\susy,\mathbb{C})$ on $H_n(L,\Sym^j(TYM))$. The action of  $H^i(\susy,\mathbb{C})$ factors through the action of $H^i(L,\mathbb{C})$

The maps $\iota_j$ in  (\ref{E:jajqud}) induced by tautological  inclusion of sheaves $\Lambda^j{\cal W}(-8-i)\rightarrow \Lambda^j(L_3)(-8-i)\subset \Sym^j(TYM)(-8-i)$. 
We claim that  $\iota_j$ are embedding. Let $p:\F\rightarrow \Q$ be a canonical projection from the full flags.

The idea is to decompose $\Lambda^j(L_3)$ into $Spin(10)$-irreducible components. Pick one of them , denoted by $A$, such that $\Lambda^j{\cal W}(-8-i)\subset A(-8-i) $. Using multiplication on sections $H^0(\Q ,{\cal O}(i))$ one can reduces the check to the case $i=0$. The Serre dual statement is a claim that  $A^*\rightarrow \Lambda^j{\cal W^*}$ induces isomorphism on global sections.  Such diagram of sheaves is a pushforward of a diagram $A^*\rightarrow \O(D)$\footnote{$D$ is a divisor} on  the full flags,  where $p_*\O(D_j)=\Lambda^j{\cal W^*}$, $p_*A^*=A^*$ On the full flags it becomes the  statement of classical Borel-Weyl theory that there is an isomorphism $A^*=H^0(\F,\O(D_j))$.
In particular the map is an embedding for $i=0$. The linear subspace $A\subset \Lambda^j(L_3)$ embeds into homology.

We can reinterpret multiplications on sections $H^0(\Q ,{\cal O}(i))$ as multiplication on elements of $H^{i}(\susy,\mathbb{C})$.

\end{proof}

\subsection{Homology of algebra $\susy$.}
In this section we compute cohomology of $\susy$ in dimension 10 with trivial coefficients.


The complex that computes Lie algebra homology is equal to a space of (super)-polynomial functions $C_{\bullet}(\susy,\mathbb{C})=\{f(u^1,\dots,u^{16},\tau^1,\dots,\tau^{10})\}$  where $u^{\alpha}$ has degree $Kdeg$  zero , $\tau^i$ -   one . The differential is given by the formula $\Gamma^{\alpha \beta}_i\xi^i\frac{\partial^2}{\partial u^{\alpha}\partial u^{\beta}}$. It is convenient to introduce a finer bigrading on $C_{\bullet}(\susy,\mathbb{C})$ by setting $deg(u^{\alpha})=1$ and $deg \xi^i=2$. The complex  $C_{\bullet}(\susy,\mathbb{C})$ then decomposes into a direct sum of complexes of elements of different degree $deg$.

\begin{equation}\label{T:dsdwcbtm}
\mbox{
\scriptsize{
$\begin{array}{|c|ccrccc}
deg   &\dots &\dots     &\dots                     &\dots     &\dots       &\dots      \\
4     &S^4(S)&\rightarrow&S^2(S)\otimes \Lambda^1(V)&\rightarrow&\Lambda^2(V)&\\
3     &S^3(S)&\rightarrow&S^1(S)\otimes \Lambda^1(V)&          &            &\\
2     &S^2(S)&\rightarrow&        \Lambda^1(V)      &          &            &\\
1     &S^1(S)&          &                          &          &            &\\
0     &\mathbb{C}&        &                          &          &            &\\\hline
      & 0    &          &1                         &           &2&Kdeg
\end{array}$
}
}
\end{equation}
\begin{proposition}\label{P:qugdgfewt}
In the table below you can find the representation content of the homology groups $H_i(\susy,\mathbb{C})$
\begin{equation}\label{T:dkeoeodd}
\mbox{
\scriptsize{
$\begin{array}{|c|c|c|c|c|c|c|c}
deg &\dots    &\dots       &\dots      &\dots      &\dots      &\dots     &\\\hline
15&\left[0,0,0,15,0\right]&\left[0,0,0,12,1\right]&\left[0,0,1,9,0\right]&\left[0,1,0,7,0\right]&\left[1,0,0,5,0\right]&\left[0,0,0,3,0\right]\\\hline
14&\left[0,0,0,14,0\right]&\left[0,0,0,11,1\right]&\left[0,0,1,8,0\right]&\left[0,1,0,6,0\right]&\left[1,0,0,4,0\right]&\left[0,0,0,2,0\right]\\\hline
13&\left[0,0,0,13,0\right]&\left[0,0,0,10,1\right]&\left[0,0,1,7,0\right]&\left[0,1,0,5,0\right]&\left[1,0,0,3,0\right]&\left[0,0,0,1,0\right]\\\hline
12&\left[0,0,0,12,0\right]&\left[0,0,0,9,1\right]&\left[0,0,1,6,0\right]&\left[0,1,0,4,0\right]&\left[1,0,0,2,0\right]&\left[0,0,0,0,0\right]\\\hline
11&\left[0,0,0,11,0\right]&\left[0,0,0,8,1\right]&\left[0,0,1,5,0\right]&\left[0,1,0,3,0\right]&\left[1,0,0,1,0\right]&          &\\\hline
10&\left[0,0,0,10,0\right]&\left[0,0,0,7,1\right]&\left[0,0,1,4,0\right]&\left[0,1,0,2,0\right]&\left[1,0,0,0,0\right]&          &\\\hline
9 &\left[0,0,0,9,0\right] &\left[0,0,0,6,1\right] &\left[0,0,1,3,0\right]&\left[0,1,0,1,0\right]&           &          &\\\hline
8 &\left[0,0,0,8,0\right] &\left[0,0,0,5,1\right] &\left[0,0,1,2,0\right]&\left[0,1,0,0,0\right]&           &          &\\\hline
7 &\left[0,0,0,7,0\right] &\left[0,0,0,4,1\right] &\left[0,0,1,1,0\right]&           &           &          &\\\hline
6 &\left[0,0,0,6,0\right] &\left[0,0,0,3,1\right] &\left[0,0,1,0,0\right]&           &           &          &\\\hline
5 &\left[0,0,0,5,0\right] &\left[0,0,0,2,1\right] &           &           &           &          &\\\hline
4 &\left[0,0,0,4,0\right] & \begin{array}{c}  \left[0,0,0,1,1\right] \\ + \\\left[0,0,0,0,0 \right] \end{array} &           &           &           &          &\\\hline
3 &\left[0,0,0,3,0\right] &\left[0,0,0,0,1\right] &           &           &           &          &\\\hline
2 &\left[0,0,0,2,0\right] &            &           &           &           &          &\\\hline
1 &\left[0,0,0,1,0\right] &            &           &           &           &          &\\\hline
0&\left[0,0,0,0,0\right]  &            &           &           &           &          &\\\hline
 &0            &1           &2          &3          &4          &5&Kdeg
\end{array}$
}
}
\end{equation}

To find the  homological dimension $dim$ of a class $\alpha\in H_i(\susy,\mathbb{C})$ of the bidegree $deg,Kdeg$, one needs to use a formula $i=deg-Kdeg$. Reviewing the table we see that there are nontrivial $\mathfrak{so}(10)$ invariant classes in degrees $3$ and $7$.
\end{proposition}
\begin{remark}
The representation content of cohomology group can be read off from the table (\ref{T:dkeoeodd}) by replacing a cell entry $[w_1,w_2,w_3,w_4,w_5]$ by $[w_1,w_2,w_3,w_5,w_4]$.
\end{remark}
\begin{proof}
Rather then giving a full proof of the  statements which is based on fairly standard technique we hint the main points.

Instead of homology, we compute cohomology. We complete the algebra \\ $\mathbb{C}[u^1,\dots,u^{16}]\otimes \Lambda[\tau_1, \dots,\tau_{10}]$, by the ideal generated by $\Gamma^i_{\alpha\beta}u^{\alpha}u^{\beta}$. Due to the grading, preserved by the differential this operation does a completion of cohomology. The completed ring can be interpreted as a ring of homogeneous  functions in a formal neighborhood of $\Q \times \mathbb{C}^{0|10}\subset \mathbf{P}^{15}\times \mathbb{C}^{0|10}$. Then we localize the complex on this neighborhood. The computation is based on spectral sequences of hypercohomology of the localized complex. 
\end{proof}
\subsection{Equivariant ordinary and cyclic homology.}\label{S:gqagbj}


For any Lie algebra $\mathfrak{g}$ and a module $N$, the cochains $C^{\bullet}(\mathfrak{g},N)$ is a module over differential graded algebra $C^{\bullet}(\mathfrak{g},\mathbb{C})$. Similarly $C_{\bullet}(\mathfrak{g},\mathbb{C})$ is a coalgebra and $C_{\bullet}(\mathfrak{g},N)$ is a comodule over it . The groups $C_{n}(\mathfrak{g},\mathbb{C})$ and $C^n(\mathfrak{g},\mathbb{C})$ are dual with a pairing $<a,b>$. Denote $\Delta(n)=\sum_i b_i\otimes n_i$-the diagonal of $n\in C_{n}(\mathfrak{g},N)$. This enables us to define an action $C^i(\mathfrak{g},\mathbb{C}) \otimes C_n(\mathfrak{g},N)\rightarrow  C_{n-i}(\mathfrak{g},N)$ by the formula $an=\sum_i<a,b_i>n_i$.

It is easy to see that the action is compatible with differentials and induces action of  $H^{i}(\mathfrak{g},\mathbb{C})$ on  $H_{n}(\mathfrak{g},N)$.


There is one more related homology theory-cyclic homology . According to \cite{Kassel} in case of universal enveloping algebras there is a different way (different than a standard which is due to Connes \cite{Connes}) to define cyclic homology which we adopt in this paper.

\begin{definition}
Suppose a complex $C$ is equipped with  two differential $b,B$, $degB=1$, $degb=-1$.  The differentials  satisfy $b^2=B^2=bB+Bb=0$.  Then $C$ is called a mixed complex.
\end{definition}
For a mixed complex $C$ define a bicomplex
\begin{center}\label{E:ydhld}
\scriptsize{
$\begin{array}{ccccc}
\dots     &          &\dots     &          &\dots\\
\downarrow b&          &\downarrow b&          &\downarrow b\\   
C^2       &\overset{B}{\leftarrow}&C^1       &\overset{B}{\leftarrow}&C^0\\
\downarrow b&          &\downarrow b&          &   \\
C^1       &\overset{B}{\leftarrow}&C^0       &          &   \\
\downarrow b&          &          &          &   \\
C^0       &          &          &          &   \\
 \end{array}$
 }
 \end{center}
Denote the total complex of the above bicomplex by $ToT(C,B,b)$. 
One can extend this to infinity to the left
\begin{center}\label{E:ydhld1}
\scriptsize{
$\begin{array}{cccccc}
&\dots     &          &\dots     &          &\dots\\
&\downarrow b&          &\downarrow b&          &\downarrow b\\  
\dots \overset{B}{\leftarrow} &C^2       &\overset{B}{\leftarrow}&C^1       &\overset{B}{\leftarrow}&C^0\\
&\downarrow b&          &\downarrow b&          &   \\
\dots \overset{B}{\leftarrow}& C^1       &\overset{B}{\leftarrow}&C^0       &          &   \\
&\downarrow b&          &          &          & \\
\dots \overset{B}{\leftarrow}& C^0       &          &          &          &   \\
\dots &        &          &          &          &   \\ \hline 
  & 0       &          &  1        &          & 2   \\
 \end{array}$
 }
 \end{center}
Denote the resulting total complex by $ToT^{per}(C,B,b)$. We allow infinite sums in $ToT^{per}(C,B,b)$.

Similarly define a bicomplex with zero entries to the right from the zero column. Denote the complex by $ToT^{-}(C,B,b)$

Fix a Lie algebra $\mathfrak{g}$. On polynomial forms $\Omega(\mathfrak{g}^*)$ of coadjoint representations of $\mathfrak{g}$ there are two differentials. The first one is the de Rham differential $d_{dR}$. 
The linear space of coadjoint representation is a Poisson manifold via Kirillov bracket $\{a,b\}$. The second differential is defined by the formula
\begin{equation}
\begin{split}
&d(a_0da_1\dots da_n)=\sum_{i=1}^n(-1)^i\{a_i,a_0\}da_1\dots \widehat{da_i}\dots da_n+\\
&+\sum_{i=1}^n(-1)^{i+j-1}a_0d\{a_i,a_j\}da_1\dots \widehat{da_i}\dots \widehat{da_j}\dots da_n
\end{split}
\end{equation}

Suppose we have an extension of Lie algebras
\begin{equation}\label{E:diwnvw}
0\rightarrow \mathfrak{l}\rightarrow \mathfrak{g}\rightarrow \mathfrak{n}\rightarrow 0
\end{equation}
Define $\Omega^{\mathfrak{n}^*}(\mathfrak{g}^*)$ as polynomial differential forms on $\mathfrak{g}^*$, which are invariant with respect to $\mathfrak{n}^*$-translations. It is easy to see that $\Omega^{\mathfrak{n}^*}(\mathfrak{g}^*)$ is closed under differentials $d_{dR}$ and  $d$
\begin{proposition}
$\Omega(\mathfrak{g}^*)$ and $\Omega^{\mathfrak{n}^*}(\mathfrak{g}^*)$ are mixed complexes, with $B=d_{dR}$, $b=d$.
\end{proposition}

\begin{definition}

$HC_n(U(\mathfrak{g}))=H_n(ToT(\Omega(\mathfrak{g}^*),d_{dR},d))$, 

$HC_n(\mathfrak{g},U(\mathfrak{l}))=H_n(ToT(\Omega^{\mathfrak{n}^*}(\mathfrak{g}^*),d_{dR},d))$.

$HC_n^{per}(U(\mathfrak{g}))=H_n(ToT^{per}(\Omega(\mathfrak{g}^*),d_{dR},d))$, 

$HC^{per}_n(\mathfrak{g},U(\mathfrak{l}))=H_n(ToT^{per}(\Omega^{\mathfrak{n}^*}(\mathfrak{g}^*),d_{dR},d))$

$HC_n^{-}(U(\mathfrak{g}))=H_n(ToT^{-}(\Omega(\mathfrak{g}^*),d_{dR},d))$, 

$HC^{-}_n(\mathfrak{g},U(\mathfrak{l}))=H_n(ToT^{-}(\Omega^{\mathfrak{n}^*}(\mathfrak{g}^*),d_{dR},d))$
\end{definition}

The following long exact sequence easily follows from definitions
\begin{equation}
\begin{split}
&\dots\rightarrow HC_{n-1}^{}(\mathfrak{g}, U(\mathfrak{l}))\rightarrow HC_{n}^{ -}(\mathfrak{g}, U(\mathfrak{l}))\rightarrow HC_{n}^{ per}(\mathfrak{g}, U(\mathfrak{l}))\rightarrow \\
&\rightarrow HC_{n-2}^{}(\mathfrak{g}, U(\mathfrak{l}))\rightarrow \dots\\
&\\
&\dots\rightarrow HC_{n+2}^{ -}(\mathfrak{g}, U(\mathfrak{l}))\rightarrow HC_{n}^{ -}(\mathfrak{g}, U(\mathfrak{l}))\rightarrow HC_{n}(\mathfrak{g}, U(\mathfrak{l}))\rightarrow \\
&\rightarrow HC_{n+1}^{}(\mathfrak{g}, U(\mathfrak{l}))\rightarrow \dots\\
&\\
&\dots\rightarrow HC_{n-1}^{ }(\mathfrak{g}, U(\mathfrak{l}))\rightarrow H_{n}(\mathfrak{g}, U(\mathfrak{l}))\rightarrow HC^{}_{n}(\mathfrak{g}, U(\mathfrak{l}))\rightarrow \\
&\rightarrow HC_{n-2}^{}(\mathfrak{g}, U(\mathfrak{l}))\rightarrow \dots\\
\end{split}
\end{equation}
Consider a complex :

\begin{equation}\label{E:DDDDDD}
\mbox{
\scriptsize{
$\begin{array}{ccccc}
\dots     &          &\dots     &          &\dots\\
\downarrow d&          &\downarrow d&          &\downarrow d\\   
\Sym^0 (\mathfrak{l})\otimes \Lambda^2 (\mathfrak{g})       &\overset{d_{dR}}{\leftarrow}&\Sym^1 (\mathfrak{l})\otimes \Lambda^1 (\mathfrak{g})       &\overset{d_{dR}}{\leftarrow}&\Sym^2 (\mathfrak{l})\otimes \Lambda^0 (\mathfrak{g})\\
\downarrow d&          &\downarrow d&          &   \\
\Sym^0 (\mathfrak{l})\otimes \Lambda^1 (\mathfrak{g})       &\overset{d_{dR}}{\leftarrow}&\Sym^1 (\mathfrak{l})\otimes \Lambda^0 (\mathfrak{g})       &          &   \\
\downarrow d&          &          &          &   \\
\Sym^0 (\mathfrak{l})\otimes \Lambda^0 (\mathfrak{g})       &          &          &          &   \\
 \end{array}$
 }
}
\end{equation}
\begin{proposition}\label{P:chdhyyye}
The cohomology of the total complex \ref{E:DDDDDD} is equal to $H_{\bullet}(\mathfrak{n},\mathbb{C})$.
\end{proposition}
\begin{proof}
To prove we use the spectral sequence of the bicomplex which $E_1$ term is equal to the cohomology of \ref{E:DDDDDD} with respect to $d_{dR}$. Choosing some linear splitting of (\ref{E:diwnvw}), we identify $\Lambda^n(\mathfrak{g})=\bigoplus_{i+j=n}\Lambda^i(\mathfrak{n})\otimes \Lambda^j(\mathfrak{l})$. The horizontal rows are equal to direct sum of homogeneous components of of the de Rham complex on $\mathfrak{l}^*$ with coefficients in $\Lambda^i(\mathfrak{n})$ for various $i$.  Due to acyclicity of the de Rham complex of $\mathfrak{l}^*$ the cohomology of $n$-th row is equal to $\Lambda^n(\mathfrak{n})$. These are located  in the first column. It is obvious that the vertical differential becomes the standard homology differential in $C_n(\mathfrak{n},\mathbb{C})=\Lambda^n(\mathfrak{n})$, described in (\ref{E:diffh}).

The spectral sequence collapses in $E_2$-term due to dimension reasons.
\end{proof}

\begin{corollary}
$HC_{n}^{per}(\mathfrak{g},U(\mathfrak{l}))=\prod_{k\in \mathbb{Z}}H_{k+n}(\mathfrak{n},\mathbb{C})$
\end{corollary}
\begin{proof}
The complex $ToT^{per}(\Omega^{\mathfrak{n}^*}(\mathfrak{g}^*)$  is a direct product of complexes \ref{E:DDDDDD}
%
\end{proof}

 $C^{\bullet}(\mathfrak{n},\mathbb{C})$ is a subalgebra of $C^{\bullet}(\mathfrak{g},\mathbb{C})$.
 We have an action of $C^i(\mathfrak{n},\mathbb{C})$ on $C_n(\mathfrak{g},U(\mathfrak{l}))$. The action is compatible with differential $d$. Moreover it commutes with $d_{dR}$.

From this we conclude that $C^{\bullet}(\mathfrak{n},\mathbb{C})$ acts on the total complex \ref{E:DDDDDD}.

The following bicomplex is a specialization of \ref{E:DDDDDD} ($\mathfrak{g}=L,\mathfrak{l}=TYM$).
\begin{equation}\label{E:DDDwwwDDD}
\mbox{
\scriptsize{
$\begin{array}{ccccc}
\dots     &          &\dots     &          &\dots\\
\downarrow d&          &\downarrow d&          &\downarrow d\\   
\Sym^0 (TYM)\otimes \Lambda^2 (L)       &\overset{d^L_{dR}}{\leftarrow}&\Sym^1 (TYM)\otimes \Lambda^1 (L)       &\overset{d^L_{dR}}{\leftarrow}&\Sym^2 (TYM)\otimes \Lambda^0 (L)\\
\downarrow d&          &\downarrow d&          &   \\
\Sym^0 (TYM)\otimes \Lambda^1 (L)       &\overset{d^L_{dR}}{\leftarrow}&\Sym^1 (TYM)\otimes \Lambda^0 (L)       &          &   \\
\downarrow d&          &          &          &   \\
\Sym^0 (TYM)\otimes \Lambda^0 (L)       &          &          &          &   \\
 \end{array}$
}
}
\end{equation}

It leads to a spectral sequence
\begin{equation}\label{E:gdafdte}
H_i(L,\Sym^j(TYM))\Rightarrow H_{i+2j}(\susy,\mathbb{C})
\end{equation}

Observe that the spaces $\Sym^j (TYM)\otimes \Lambda^i (L)$ for fixed $j$ form columns of \ref{E:DDDwwwDDD}. It makes sense to talk about cocycles of $C_i(L, \Sym^j (TYM))$ as of elements of \ref{E:DDDwwwDDD}.

\begin{proposition}\label{P:wsidcnxs}
Every element in the image of the maps $\iota_s,s=1\dots 5$ in (\ref{E:jajqud})  for $i\geq 4 $ can be modified to a cocycle of the total complex \ref{E:DDDwwwDDD}.
\end{proposition}
\begin{proof}
The maps $d$ and $d_{dR}$ are $Spin(10)$-equivariant. We can utilize this to solve ``tic-tac'' process: pick an element $a=a_0$  whose class is  from the image of $\iota_s$, moreover it belongs to an irreducible $Spin(10)$ subrepresentation of $\Sym^j (TYM)\otimes \Lambda^i (L)$. The representation must be one of in (\ref{E:jajqud}). The element $d_{dR}a$ belongs an irreducible representation of the same isotopic type. From proposition (\ref{P:qideesa}) we know that it must be homologous to zero. We find $a_1$ again in irreducible subrepresentation of  $\Sym^{j-1} (TYM)\otimes \Lambda^{i+1} (L)$ such that $da_1=d_{dR}a$. We continue this process until we end up in $\Sym^{0} (TYM)\otimes \Lambda^{i+j} (L)$. Let $a_0=a$. The sequence of elements $(a_0,a_1.\dots,a_j)$ is a cocycle of \ref{E:DDDwwwDDD}.
\end{proof}

\begin{lemma}\label{L:wsdsfsc}
The maps $\iota_j, 1\leq j \leq 5$ in (\ref{E:jajqud}) are embeddings  for $i\geq0$ .
\end{lemma}
\begin{proof}
It is given in section (\ref{S:localiz}).
\end{proof}

Denote the image of $\iota_j$ in degree $i$ by $A_j^i$
\begin{lemma}\label{L:twxctw}
There is a  cocycle representing  any class in $\Im \iota_j$ $i\geq 0$ which can be lifted to nontrivial cocycles of \ref{E:DDDwwwDDD}.
\end{lemma}
\begin{proof}
The proof is a combination of proofs of proposition (\ref{P:wsidcnxs}) and lemma (\ref{L:wsdsfsc}).

Indeed, as in a proof of proposition (\ref{P:wsidcnxs}), lift a cocycle $a$ to a cocycle $a=a_0,\dots,a_i$. Apply multiplication on $\lambda^{\alpha_s}$ several times, until $\lambda^{\alpha_1}\dots \lambda^{\alpha_i} a_0$ lands  in zero chains $\Sym^j(TYM)$. As we know from lemma (\ref{L:wsdsfsc}), this element projects nontrivially into  $H_0(L,\Sym^j(TYM))$. Thus the cochain $\lambda^{\alpha_1}\dots \lambda^{\alpha_i}(a_0,\dots,a_i)$ has a nontrivial class in cohomology. From this we conclude that so does the class $(a_0,\dots,a_i)$.

\end{proof}

The cohomology of the total complex \ref{E:DDDwwwDDD} is easy to compute.
\begin{remark}
According to proposition (\ref{P:chdhyyye})  applied to $\mathfrak{g}=L$, $\mathfrak{l}=TYM$, the cohomology of \ref{E:DDDwwwDDD} is equal to $H_{\bullet}(\susy,\mathbb{C})$.
\end{remark}
According to lemma (\ref{L:twxctw}) every class in $\Im \iota_j$ gives a nontrivial contribution to $H_{\bullet}(\susy,\mathbb{C})$.
We see that the image of the maps $\iota_j$ cover almost all cohomology of $\ref{E:DDDwwwDDD}$ (see the table (\ref{T:dsdwcbtm}) in appendix).

The exceptional representations in homology $H_{\bullet}(\susy,\mathbb{C})$ not covered by the above construction are 
\begin{equation}
\begin{split}
&[0,0,0,2,0]\subset H_9(\susy,\mathbb{C}), 
[0,0,0,1,0]\subset H_8(\susy,\mathbb{C}), \\
&[0,0,0,0,0]\subset H_7(\susy,\mathbb{C}), 
[1,0,0,1,0]\subset H_7(\susy,\mathbb{C}), \\
&[1,0,0,0,0]\subset H_6(\susy,\mathbb{C}), 
[0,1,0,0,0]\subset H_5(\susy,\mathbb{C}), \\
&[0,0,0,0,0]\subset H_3(\susy,\mathbb{C})
\end{split}
\end{equation}

To do this we need to prove the following 
\begin{proposition}\label{P;mxswess}
$H_3(L,\Sym^j(TYM))=A_j^3$, $H_2(L,\Sym^j(TYM))=A_j^2$
\end{proposition}
\begin{proof}
The proof is given in section (\ref{S:oodsasfd}).
\end{proof}
\begin{proposition}\label{P:xiuwbx}
The spectral sequence (\ref{E:gdafdte}) collapses in $E_2$-term.
\end{proposition}
\begin{proof}
We already know that classes in $A_j^i$ survive to $E_{\infty}$. Due to proposition (\ref{P;mxswess}) and dimension reasons  the spectral sequence collapses .
\end{proof}
\begin{corollary}\label{C:gfdcbeyh}
The differential 
\begin{equation}\label{E:vhbdc}
d_{dR}:H_1(L,\Sym^j(TYM))\rightarrow H_{2}(L,\Sym^{j-1}(TYM))
\end{equation}
 is zero for $j\geq 0$.
\end{corollary}
\begin{proof}
According to proposition (\ref{P;mxswess}) $H_2(L,\Sym^j(TYM))=A_j^2$. This groups survives to $E_{\infty}$, thus the map (\ref{E:vhbdc}) is trivial.

\end{proof}
\begin{corollary}\label{C:gfddxfc}
There are linear subspaces
\begin{equation}\label{E:qjxhwyx}
\begin{split}
&[0,0,0,1,0]\subset H_{0,13}(L,\Sym^4(TYM))\\
&[1,0,0,0,0]\subset H_{0,10}(L,\Sym^3(TYM))\\
&\\
&\mathbb{C}_{1,4}\subset H_{1,4}(L,TYM)\\
&[0,1,0,0,0]\subset H_{1,8}(L,\Sym^2(TYM))\\
&\mathbb{C}_{1,12}\subset H_{1,12}(L,\Sym^3(TYM))\\
&[1,0,0,1,0]\subset H_{1,11}(L,\Sym^3(TYM))\\
&[0,0,0,2,0]\subset H_{1,14}(L,\Sym^4(TYM))
\end{split}
\end{equation}
\begin{equation}\label{E:dfdsfyh}
\begin{split}
&\mbox{The space } \mathbb{C}_{1,4} \mbox{ is spanned by } \gamma_{1,4}=\sum_{\alpha}\lambda^{*}_{\alpha}\chi^{\alpha}\\
& \mbox{ the space } \mathbb{C}_{1,12} \mbox{  is spanned by } \gamma_{1,12}=\Gamma_{\alpha}^{\beta[i_1,\dots,i_4]}\lambda^{*}_{\alpha}\otimes \chi^{\alpha}\circ F_{i_1i_2}\circ F_{i_3i_4}
\end{split}
\end{equation}
 Where $F_{ij}=[v_i,v_j]$and $\circ$ is a graded symmetric product.

The spaces (\ref{E:qjxhwyx}) belong to the kernel of $d_{dR}$. Denote a direct sum of the above subspaces of  $ H_{i}(L,\Sym^j(TYM))$ by $B_j^i$

Fix $n$. The direct sum  $B_j^{n-j}+A_j^{n-j}$   isomorphically project onto $j$-th cohomology of the complex $(H_{n-j}(L,\Sym^j(TYM)),d_{dR})$.
\end{corollary}
\begin{proof}
Simple use of propositions (\ref{P:xiuwbx}, \ref{P;mxswess}, \ref{P:qugdgfewt})
\end{proof}

\begin{proposition}
The maps 
\begin{equation}\label{E:dvfsdhh}
\pi_j^iH^{j+i,2j-i}(L,\Sym^{j}(TYM))\rightarrow H^0(\Q ,\Lambda^j(\W)(j+i))
\end{equation} 
$i,j\geq 0$ defined in (\ref{P:azbush}) are surjections.
\end{proposition}
\begin{proof}
It is easy to see that the map $\pi_j^i$ maps cocycle $\Gamma^s_{\alpha\beta}\lambda^{\alpha}\chi^{\beta}$ to a generator $v_s\in  H^0(\Q ,\Lambda^1(\W)(1))$ $s=1\dots,10$. The products of $v_s$ and $\lambda^{\alpha}\in H^0(\Q ,{\cal O}(1))$ generate $ H^0(\Q ,\Lambda^j(\W)(j+i))$. We interpret $\lambda^{\alpha}$ as elements of $H^1(L,\mathbb{C})$. 

We prove proposition using that $\pi$ is an algebra homomorphism, the target has no zero divisors as an algebra over $S$  and $\pi_j^i$ is an isomorphism for $i\geq 4$
\end{proof}

\begin{proposition}
The following maps defined in proposition (\ref{P:azbush}) are surjections.
\begin{equation}\label{E:jauuyd}
\begin{split}
&H^{3,12}(L,\Sym^2(TYM))\rightarrow  H^9(\Q ,\Lambda^2(\W)(-6))=\mathbb{C}\\
&H^{2,8}(L,\Sym^3(TYM))\rightarrow  H^1(\Q ,\Lambda^3(\W)(1))=\mathbb{C}\\
&H^{3,16}(L,\Sym^3(TYM))\rightarrow  H^{10}(\Q ,\Lambda^3(\W)(-7))=[0,1,0,0,0]\\
&H^{4,18}(L,\Sym^4(TYM))\rightarrow  H^{10}(\Q ,\Lambda^4(\W)(-6))=[1,0,0,0,0]\\
&H^{3,19}(L,\Sym^4(TYM))\rightarrow  H^{10}(\Q ,\Lambda^4(\W)(-7))=[1,0,0,1,0]\\
&H^{5,20}(L,\Sym^4(TYM))\rightarrow  H^{10}(\Q ,\Lambda^5(\W)(-5))=\mathbb{C}\\
&H^{4,21}(L,\Sym^5(TYM))\rightarrow  H^{10}(\Q ,\Lambda^5(\W)(-6))=[0,0,0,1,0]\\
&H^{3,22}(L,\Sym^5(TYM))\rightarrow  H^{10}(\Q ,\Lambda^5(\W)(-7))=[0,0,0,2,0]
\end{split}
\end{equation}
\end{proposition}
\begin{proof}In long exact sequence (\ref{E:ywers}) all linear spaces $H^{i}(\Q ,\Lambda^j(\W)(k))$ from (\ref{E:jauuyd}) are mapped to a zero space.
\end{proof}

Consider a complex \ref{E:DDDDDD} for a pair $TYM\subset YM$. 
\begin{proposition}\label{P:ayvdc}
The spectral sequence of bicomplex \ref{E:DDDDDD} for a pair $TYM\subset YM$ has the only nontrivial differential in  $E_2$-term, which maps a class of Lagrangian in $H_0(YM,\Sym^2(TYM))$ to the generator of $H_3(YM,\mathbb{C})$
\end{proposition}
\begin{proof}
We leave the proof of this proposition as an exercise for the  reader. Most of the differentials are zero because of vanishing of $H_3(YM,\Sym^k(TYM))$.
\end{proof}
\begin{proposition}
The spectral sequence of proposition (\ref{P:ayvdc}) splits into exact sequences:
\begin{equation}\label{E:gfdfdfgj}
\begin{split}
&0\rightarrow \Lambda^{2i}(V)\rightarrow H_0(YM,\Sym^i(TYM))\overset{d^{YM}_{dR}}{\rightarrow}H_1(YM,\Sym^{i-1}(TYM))\overset{\nu}{\rightarrow}\\ & \overset{\nu}{\rightarrow} \Lambda^{2i-1}(V)\rightarrow 0 \quad i>3\\
&0\rightarrow \Lambda^{6}(V)\rightarrow H_0(YM,\Sym^3(TYM))\overset{d^{YM}_{dR}}{\rightarrow}H_1(YM,\Sym^{2}(TYM))\overset{\nu}{\rightarrow}\\ & \overset{\nu}{\rightarrow} \Lambda^{5}(V)+H_2(YM,TYM)\rightarrow 0 \\
&0\rightarrow \Lambda^{4}(V)\rightarrow H_0(YM,\Sym^2(TYM))/\mathbb{C}\overset{d^{YM}_{dR}}{\rightarrow}H_1(YM,\Sym^{1}(TYM))\overset{\nu}{\rightarrow}\\ & \overset{\nu}{\rightarrow} \Lambda^{3}(V)+H_2(YM,\mathbb{C})\rightarrow 0 \\
&0\rightarrow \Lambda^{2}(V)\rightarrow H_0(YM,TYM)\overset{d^{YM}_{dR}}{\rightarrow}H_1(YM,\mathbb{C})\overset{\nu}{\rightarrow}\\ & \overset{\nu}{\rightarrow} \Lambda^{1}(V)\rightarrow 0 \\
&
\end{split}
\end{equation}
\end{proposition}

\begin{remark}
The map $d^{YM}_{dR}$ corresponds to taking equations of motion of a Lagrangian. The later is an element of zero homology group. The kernel of the map $d^{YM}_{dR}$ are topological term of the Lagrangian.
\end{remark}
\begin{remark}
We can say that the elements of $H_2(YM,\Sym(TYM))$ which do not map to zero under $\nu$ are nonlagrangian deformations of equations of motion.
\end{remark}
\begin{remark}\label{R:fdsdsaq}
By proposition (\ref{P:hygdscv}) $res\delta d^L_{dR}f$, $f\in H_0(L,\Sym(TYM))$ automatically belongs to the image of $d^{YM}_{dR}$ and hence is a Lagrangian deformation. The cocycle $c_{2,8}\in H^{2,8}(L,\Sym^3(TYM))$ has Poincare dual in $H_{1,16}(YM,\Sym(TYM))$. The later element has image under $\nu$ equal to zero in (\ref{E:gfdfdfgj}) due to degree considerations. The same arguments applies to cocycles $\delta\gamma_{1,4},\delta\gamma_{1,12}$.
\end{remark}

\subsection{On the structure of the connecting differential  $\delta$ from (\ref{E:delta})}

The map $\delta$ plays an important role in our construction of infinitesimal deformations. In this section we make a preliminary study of $\delta$ and identify it with a cocycle in $HH^{\bullet}(S,S\otimes S)$.
Let $N$ be an $L$-module.
The table below represents $E_1$ term of a spectral sequence of a bicomplex  $E^{i,j}_1\Rightarrow H^{i+j}(\Omega\N(l))$. See definition (\ref{D:fgysvbc})  for explanation of notations.
\begin{center}\label{E:yuf}
\scriptsize{
$\begin{array}{|c|ccccccccccccc}
&&N_{-l-10}&&N_{-l-9}&&&&&&&&\\
10 & \dots & \otimes  & \rightarrow &\otimes  & \rightarrow &N_{-l-8} &\dots & 0 &  & 0 &  &0&\dots\\
&&{\cal S}^*_{2}&&{\cal S}^*_{1}&&&&&&&&\\
9 &\dots &0&&0&&0&\dots&0&&0&&0&\dots \\
\dots & \dots&\dots&&\dots&&\dots&\dots&\dots&&\dots&&\dots\\
1&\dots&0&&0&&0&\dots&0&&0&&0&\dots \\
&&&&&&&&&&N_{-l+1}&&N_{-l+2}\\
0&\dots&0&&0&&0&\dots&N_{-l}&\rightarrow &\otimes&\rightarrow &\otimes&\dots\\
&&&&&&&&&&{\cal S}_{1}&&{\cal S}_{2}\\\hline
&&-l-10&&-l-9&&-l-8&&-l&&-l+1&&-l+2&

\end{array}$
}
\end{center}
%

Let us describe the connecting differential of (\ref{E:yuf}) in analytic terms. Fix $Spin(10,\mathbb{R})$-invariant Kahler metric on $\Q $ and on ${\cal O}(1)$
Let $p_k:\Omega^{0.k}(l) \rightarrow H^k(\Q ,\O(l)) $ be the orthogonal projection onto cohomology. We have $p_k=0$ for $k \neq 0,10 $. Let us set $p=\bigoplus_{k=0}^{10} p_k$. 
One can choose $SO(10)$-equivariant homotopy $L:\Omega^{0.k}(l)\rightarrow \Omega^{0.k-1}(l)$ which satisfies the following set of properties:$L^2=0$, $\{\dbar,L\}=Id-p$, where $Id$ is the identity transformation.

Suppose $a=\sum a_i\otimes\omega_i\in N\otimes S^*$ is a representative of a cohomology class of one of cohomology groups in the tenth row  of table (\ref{E:yuf}). Let 
$e=\sum_{\alpha} \theta_{\alpha}\lambda^{\alpha}$. The element $b_1=\sum  \theta_{\alpha}a_i\otimes\lambda^{\alpha} \omega_i$ is $\dbar$-coboundary. Define an element $c_1=\sum \theta_{\alpha}a_i\otimes L(\lambda^{\alpha} \omega_i)$. By construction $\dbar c=b_1$. Iterating this construction, we get  an element
\begin{equation}\label{E:pjh}
 \theta_{\alpha_1}\theta_{\alpha_2}\dots \theta_{\alpha_{11}}a_i\dots\otimes\lambda^{\alpha_1} L(\lambda^{\alpha_2} \dots L (\lambda^{\alpha_{11}}  \omega_i))\dots))
\end{equation}
, which is  $\delta$ differential of element $a$.

There is an algebraic  description of the differential $\delta$. 

To give it  we need to make a digression. Let $A$ be an algebra $N',N''$ are two modules. We can define  groups $Ext^i_A(N',N''), i>0 $ following Yoneda.  Consider an acyclic complex of modules $$0\rightarrow N'' \rightarrow P_1 \rightarrow \dots \rightarrow P_{i}\rightarrow N' \rightarrow 0$$ These exact sequences form a  semigroup . There is an obvious notion of a equivalence of such sequences and  a notion of a "trivial" sequences. After factorization with respect to equivalence relation and after killing all trivial elements we get groups $Ext^i_A(N',N'')$ (see \cite{McL} for details).

We utilize  the  Dolbeault complex $\bigoplus_{l \in \mathbb{ Z}} \Omega^{0,\bullet}(l)$ to construct an element in $Ext^{11,8}_{S}(S^*,S)$.  The Dolbeault differential $\bar{\partial}$ is linear with respect to multiplication on elements of $S\subset \bigoplus_{l \in \mathbb{ Z}} O(l)\otimes \Omega^{0,0}$. It means that we can interpret a complex 
\begin{equation}\label{E:gadsgj}
0 \rightarrow S \rightarrow \bigoplus_{l \in \mathbb{ Z}} \Omega^{0,0}(l) \rightarrow \dots \rightarrow \bigoplus_{l \in \mathbb{ Z}} \Omega^{0,10}(l)\rightarrow {\cal S}^* \rightarrow 0
\end{equation}
as an element of $Ext_S^{11,8}(S^*,S)$.

According to \cite{CE} for any algebra $A$ and two left modules $M,N$ there is an isomorphism $Ext^n_A(M,N)=HH^n(A,N\otimes M^*)$. We conclude $Ext^n_S(S^*,S)=HH^n(S,S\otimes S)$. The later group can be computed via Koszul resolution:
\begin{proposition}\label{P:veyyc}
The cohomology $HH^n(S,S\otimes S)$ can be computed as cohomology of the complex $U(L)\otimes S\otimes S$. The differential is defined for homogeneous elements by the formula
\begin{equation}
d(a\otimes b \otimes c)=\left(\theta_{\alpha}a\otimes \lambda^{\alpha}b \otimes c-(-1)^{\tilde a}a\theta_{\alpha}\otimes b \otimes \lambda^{\alpha}c\right)
\end{equation}
\end{proposition}
\begin{proof}
is similar to (\ref{P:tqydxc}) and (\ref{P:iqwwst}) 
\end{proof}
\begin{proposition}
The cohomology $HH^n(S,S\otimes S)$ is equal to $0$ for $n\neq 11$ and $S$ for $n=11$
\end{proposition}
\begin{proof}
We need to start with a remark that this statement is true for coordinate ring of any smooth affine variety. In non smooth case this statement is typically not correct, which makes this proposition a bit surprising.

We do computations with  $U(L)\otimes S\otimes S$ as it was explained in (\ref{P:veyyc}). The statement of proposition easily follows from the following fact:
\begin{lemma}
The cohomology of the complex $U(L)\otimes S$ with differential equal to left multiplication on $\lambda^{\alpha}\theta_{\alpha}$ is equal to $\mathbb{C}$. The nontrivial cocycle has bidegree $(11,3)$.
\end{lemma}
\begin{proof}
As usual, we use localization arguments. Fix $\theta \in L_1$, such that $\theta^2=0$. The cohomology of $U(L)$ with differential-on $\theta$ are trivial(use a spectral sequence associated with PBW filtration to prove this statement).

The spectral sequence of the hypercohomology of the localized complex $({\cal U(L)}, d)$ for  $(U(L)\otimes S,d)$ converges to zero and collapses in $E_11$. The only nontrivial higher differential $d_10$ in $E_3=E_10$ establishes isomorphism between cohomology $U(L)\otimes S^*$ and $U(L)\otimes S$

The algebra $S$ is Koszul. Thus by definition the complex $U(L)\otimes S^*$ has the only nontrivial cohomology in bidegree $(0,0)$. The image of this cohomology class in $U(L)\otimes S$ is the one with bidegree $(11,3)$.
\end{proof}

\end{proof}

Let $N$ be an $L$-module. There is a sequence of maps 
\begin{equation}
H_0(YM,N)\overset{i}\rightarrow H_0(L,N)\overset{\delta}\rightarrow H^3(L,N)\overset{res}\rightarrow H^3(YM,N)\overset{P}\rightarrow H_0(YM,N)
\end{equation}
The map  $i$ is a map of zero homology of subalgebra on zero homology of algebra,  $\delta$ is the differential (\ref{E:pjh}), $res$ is the restriction map from cohomology of algebra to cohomology of subalgebra, $P$ is the Poincare isomorphism for the  algebra $YM$.

Denote the composed map by $\psi_0$.

\begin{proposition}
Denote $x=\epsilon^{\alpha_1\dots\alpha_{16}}\theta_{\alpha_1}\dots\theta_{\alpha_{16}}$ an element in $U(L)$.
The map $\psi_0$ is defined by the formula 
\begin{equation}\label{E:ddshc}
\psi_0(n)=xm
\end{equation}
\end{proposition}
\begin{proof}
If we change the map $xm$ by adding to $x$ an element $\tilde x$ of the same degree but lower order in PBW filtration the map $xm+\tilde xm$ 
is still equal to $xm$. 

Denote $YM(N)=\{n\in N|n=\sum_il_in_i,l_i\in YM, n_i\in N\}$. It is easy to see that on the level of elements of $N$, $\tilde xn\in YM(N)$, but by definition $H_0(YM,N)=N/YM(N)$

The map $\psi_0$ is completely determined by cocycle (\ref{E:gadsgj}), which we interpret as an element of $H^3(L,U(L))$. The Lie algebra $L$ acts on $U(L)$ by left multiplication. 

We will reinterpret this cocycle through spectral sequence of extension $YM\subset L$. The $E_1^{ij}$-term is equal to $C^i(L/YM,H^j(YM,U(L)))$. 
It is easy to see that $U(L)$ is free as a $U(YM)$-module under left multiplication(in fact this is true for any Lie algebra and subalgebra). We conclude that $H^j(YM,U(L))=0$, $j\neq 3$ and $\mathbb{C}\underset{U(YM)}{\otimes}U(L)=\Lambda(L_1)$ for $j=3$. The $E_2^{i,3}$ is equal to cohomology of Koszul complex $H^{\bullet}(L_1,\Lambda(L_1))=H^{\bullet}(\Sym(L_1^*)\otimes \Lambda(L_1))$. The cohomology is one-dimensional and are represented by cocycle $\epsilon^{\alpha_1\dots\alpha_{16}}\theta_{\alpha_1}\dots\theta_{\alpha_{16}}\in \Lambda^{16}(L_1)$. A lift of this class to a cocycle in $C^3(YM,U(L))$ and identification with  $C_0(YM,U(L))$ involves ambiguities (choses in lower order of PBW-filtration), which are do not affect the final map as we showed above.
\end{proof}

There is a minor generalization of the map $\psi$. There  are compositions of maps 

\begin{equation}
 H_i(YM,N)\overset{\iota}\rightarrow H_i(L,N)\overset{\delta}\rightarrow H^{3-i}(L,N)\overset{res}\rightarrow H^{3-i}(YM,N)\overset{P}\rightarrow H_i(YM,N)
\end{equation}
We denote them by $\psi_i$.
\begin{proposition}\label{P:hygdscv}
The maps $\psi_iH_i(YM,\Sym(TYM))\rightarrow H_i(YM,\Sym(TYM))$ commute with differential $d^{YM}_{dR}$.
\end{proposition}
\begin{proof}
Any derivation of $YM$ which preserves $TYM$ acts on $H_i(YM,\Sym(TYM))$ due to functionality. Due to the same functionality the action of derivations is compatible with $d^{YM}_{dR}$. It implies that $d^{YM}_{dR}$ is compatible with the action of $x=\epsilon^{\alpha_1\dots\alpha_{16}}\theta_{\alpha_1}\dots\theta_{\alpha_{16}}$.
\end{proof}

\subsection{Computation of $H^0(L,\Sym^j(TYM))$, $H^1(L,\Sym^j(TYM))$}\label{S:oodsasfd}

According to long exact sequence (\ref{E:ywers}) and proposition (\ref{L:wsdsfsc}), we have an isomorphism $H_3(L,\Sym^j(TYM))/A_j^3=H^3(L,\Sym^j(TYM))$. $H_2(L,\Sym^j(TYM))/A_j^3=H^1(L,\Sym^j(TYM))$.

The present section content is a proof of the following  lemma 

\begin{lemma}
$H^1(L,\Sym^j(TYM))=0$, $j\geq 2$ and $H^1(L,TYM)=[1,0,0,0,0]$ and $H^0(L,\Sym^j(TYM))=0, j\geq 1$, elements $H^1(L,TYM)$ have degree two.
\end{lemma}

For the proof we adopt a method  developed in \cite{M3} where we treated a similar problem in a more simple context of pure Yang-Mills theory.

There is a short exact sequence of algebras 
where $L_1$ is an abelian Lie algebra in degree one.

For an  estimate of $H^i(L,\Sym^k(TYM))$ we can use a Serre-Hochschild spectral sequence 
(\ref{E:dhupksquy}).
If we manage to prove that $H^1(YM,\Sym^k(TYM))=0, k\geq 2$, $H^1(YM,TYM)=[1,0,0,0,0]+[1,0,0,1,0]$, $H^0(YM,\Sym^k(TYM))=0$, $k\geq 1$ we shall be done.

The text below follows very close to \cite{M3}. We omit the proofs which are close to their bosonic counterparts.

We proceed as in \cite{M3} replacing $\Sym(TYM)$ by $U(TYM)$. The universal enveloping of $TYM$ is a free algebra generated by $\mathbb{Z}_2$ graded linear space $M$. Define a filtration by powers of augmentation ideal $I\subset U(TYM)$.  The adjoint action of $YM$ on $\bigoplus_{i\geq 0}Gr_iI^i/I^{i-1}=\bigoplus_{i\geq 0}M^{\otimes i}$ factors through abelenization $Ab(YM)=L_2+L_3$. 
The component $L_3$ of $Ab(YM)$ acts on $\bigoplus_{i\geq 0}M^{\otimes i}$ trivially.

Let us describe the linear space $M$, which is automatically $U(L_2)=\Sym(V)$-module. It consists of two components $M_0$ and $M_1$ (see \cite{MSch2}). They admit a geometric interpretation.

Consider a nonsingular quadric $\X \subset \mathbf{P}^{9}$ defined by equation $q=0$ where 
\begin{equation}
q=x_i^2 \in \Sym(V)
\end{equation}
The polarization of defining equation is the bilinear form, used in definition of algebra $YM$. Let $T$ be the tangent bundle to $X$ . Define  $M_0=\bigoplus_{i\geq 0}H^0(\X,T(i))$.

The quadric is a homogeneous space of $Spin(10)$. The Levi subgroup of the stabilizer of a point $x\in \X$ is $\mathbb{C}^*\times Spin(8)$. The group $Spin(8)$ acts on the fiber via fundamental $8$-dimensional representation. 

There is a famous triality for the group $Spin(8)$. It collects all representations of $Spin(8)$ in orbits  $S_3$ action. 

The orbit of the  defining representation $v$ of  $Spin(8)$ contains $s^+$ and $s^-$- the spinor representations. They also have dimension eight.

To define $M_1$ we take $s^+$ and induce a holomorphic vector bundle on $\X$, which we denote by ${\cal s}^+$. Then  $M_1=\bigoplus_{i\geq 0} H^{0}(X,{\cal s}^+(i))$.

Fix a $YM$-module $N$. The following is a complex constructed from a free $U(YM)$-resolution of trivial $YM$-module $\mathbb{C}$. The complex computes $H^i(YM,N)=H_{3-i}(YM,N)$
\begin{equation}\label{E:dft}
N\overset{d_1}{\rightarrow} N\otimes (V+S^*)\overset{d_2}{\rightarrow} N\otimes (V+ S)\overset{d_3}{\rightarrow} N 
\end{equation}

\begin{align}
&d_1(f\otimes c)=v_if\otimes v_i^{*}+(-1)^{\tilde f}\chi^{\alpha}f\otimes \chi^{*}_{\alpha} \notag\\
&d_2(f\otimes v^{*}_k)=\left(v_iv_if\otimes v_k+v_iv_kf\otimes v_i-2v_kv_if\otimes v_i\right)+\notag\\
&+(-1)^{\tilde f}\Gamma_{\alpha\beta}^k\chi^{\alpha}f\otimes \chi^{\beta}\notag\\
&d_2(h\otimes \chi^{*}_{\alpha})=-\Gamma_{\alpha\beta}^i\left(v_ih\otimes\chi^{\beta}-(-1)^{\tilde h}\chi^{\beta}h\otimes v_i\right)\notag\\
&d_3(f\otimes v_i+h\otimes \chi_{\alpha})=v_if+(-1)^{\tilde{h}}\chi_{\alpha}h\notag
\end{align}
%
In case of module $M^{\otimes n}$ the complex (\ref{E:dft}) splits into sum of two complexes:
\begin{align}
&<c^*>\otimes M^{\otimes n} \overset{d_1^{YM}}{\rightarrow} V\otimes M^{\otimes n} \overset{d_2^{YM}}{\rightarrow} V\otimes M^{\otimes n} \overset{d_{3}^{YM}}{\rightarrow} <c>\otimes  M^{\otimes n}\label{E:ydhldi1e}\\
&\quad\quad\quad\quad\quad\quad\quad\quad\quad  S^*\otimes M^{\otimes n} \overset{d_2^D}{\rightarrow} S\otimes M^{\otimes n}  \label{E:ydhldi1e2}
\end{align}
since the action of odd generators of $YM$  on $M$ is trivial. We denote the above direct sum by $SC( M^{\otimes n})$ 


Let $N$ be a $\Sym(V)$-module. Consider a complex $N\otimes \Lambda(V)$. The linear space  $V\subset \Lambda(V)$ has a basis $\varsigma_1,\dots,\varsigma_{10}$ in degree one. A linear subspace  $V\subset \Sym(V)$ has  a basis $x_1\dots,x_{10}$  in degree two. The differential $d$ in the complex $N\otimes \Lambda(V)$ is defined by the formula $d(n\otimes \varsigma_{i})=x_in$ on generators  and extended by the Leibniz rule.
\begin{proposition}
There is a long exact sequence
\begin{equation}\label{E:focnmd}
\begin{split}
&0\rightarrow H_3(V,M^{\otimes j})\overset{S_j}{\rightarrow} H_1(V,M^{\otimes (j+1)})\overset{B_j}{\rightarrow} H_2(YM,M^{\otimes j})\overset{I_j}{\rightarrow}\\
& \rightarrow H_2(V,M^{\otimes j})\overset{S_j}{\rightarrow} H_0(V,M^{\otimes (j+1)})\overset{B_j}{\rightarrow} H_1(YM,M^{\otimes j})\overset{I_j}{\rightarrow} H_1(V,M^{\otimes j})\rightarrow 0 
\end{split}
\end{equation}

and isomorphisms
\begin{equation}
H_3(YM,M^{\otimes j})=0 \quad j \geq 1
\end{equation}
\begin{equation}\label{E:bnxgsetygf}
H_0(YM,M^{\otimes j})=H_0(V,M^{\otimes j})
\end{equation}
\begin{equation}\label{E:bsye}
H_s(V,M^{\otimes j})=\Lambda^{2j+s}[V]\quad s\geq 2, \mbox{ except } s=2,j=1
\end{equation}
For $j=1$ we have
\begin{equation}\label{E:shgscsh}
\begin{split}
&0\rightarrow \Lambda ^4V\rightarrow H_2(V,M_0)\rightarrow \mathbb {C}\rightarrow 0\\
&0\rightarrow \Lambda ^3V\rightarrow H_1(V,M_0)\rightarrow V\rightarrow 0\quad H_1(V,M_1)=S\\
&\quad \Lambda ^2V=H_0(V,M_0)\quad H_1(V,M_1)=S^*
\end{split}
\end{equation}
There is $\Sym(V)$-linear map 
\begin{equation}
\delta^c:H_{i}(YM,M^{\otimes j})\rightarrow H_{i+1}(YM,M^{\otimes j-1})
\end{equation}

Denote composition $B_{j-1}\circ I_j=\delta^c_j$. Then $\delta^c_{j-1}\circ \delta^c_j=0$

\end{proposition}
\begin{proof}
The proof repeats the proof of proposition 10 in \cite{M3}. 
To compute $H_i(V,M_1)$ we used a free $\Sym(V)$-resolution of length two from \cite{MSch2}:
\begin{equation}
M_1\leftarrow S\otimes \Sym(V)\leftarrow S^*\otimes \Sym(V)\leftarrow 0
\end{equation}
Suppose $s_{\alpha}$ is a basis $S$ and $s^{\alpha}$ dial basis of $S^*$. Then $ds^{\alpha}=\Gamma^{\alpha\beta}_ix^{i}s_{\alpha}$.

\end{proof}

%
%
\begin{corollary}\label{C:qasaw2}
$H_i(V,M^{\otimes n})=0$ if $n\geq 2$ and $i=9,10$.
\end{corollary}
%
\begin{proposition}
For $n\geq 2$ the operator of multiplication on $q$ in $M^{\otimes n}$ has no kernel.
\end{proposition}
\begin{proof}
The same as a proof of proposition 17 in \cite{M3}.
\end{proof}
\begin{proposition}\label{P:qsfwqcc}
For $n\geq 2$ the first cohomology of the complex $SC(M^{\otimes n})$ equal to $H^1(YM,M^{\otimes n})$ are trivial.
\end{proposition}
\begin{proof}
The complex $SC(M^{\otimes n})$ splits into sum of two: the complex \ref{E:ydhldi1e}   $C(M^{\otimes n})$ (defined in \cite{M3} equation 17) and the  complex \ref{E:ydhldi1e2}.

The first cohomology of the complex $C(M^{\otimes n})$ is equal to the group $H(M^{\otimes n})$ (see definition 13 \cite{M3}). Since $H_i(V,M^{\otimes n})=0$, $i=9,10$ and operator of multiplication on $q$ in $M^{\otimes n}$ has no kernel then by proposition 15 \cite{M3} $H(M^{\otimes n})=0$

The complex \ref{E:ydhldi1e2}  admits a homotopy $M^{\otimes n}\otimes S^* \overset{H}\leftarrow M^{\otimes n}\otimes S$, defined by the formula $H(m\otimes s_{\alpha})=\Gamma_{\alpha\beta}^ix_im\otimes s^{\beta}$. The composition $Hd_D$ is proportional to an operator of multiplication on $q$, which has  zero kernel in  $M^{\otimes n}$, $n\geq 2$. It implies that $d_D$ has no kernel. 

Thus the complex $SC(M^{\otimes n})$ has vanishing first cohomology.
\end{proof}

\begin{proposition}\label{P:udhgewgfa}
The zero cohomology of $SC(M^{\otimes n})$, equal to $H^0(YM,M^{\otimes n})$ are zero for $n\geq 1$.
\end{proposition}
\begin{proof}
Zero cohomology of the complex is equal to $H_{10}(V, M^{\otimes n})$. The later group is zero by corollary \ref{C:qasaw2}.
\end{proof}

\begin{proposition}\label{P:vsbfh}
Denote 
\begin{equation}
\begin{split}
&A=\bigoplus_{i\geq 0}\left([i,2,0,0,0]+[i+2,0,0,0,0]+[i+1,0,1,0,0]\right)\\
&B=\bigoplus_{i\geq 0}\left([i,1,0,0,1]+[i+1,0,0,1,0]\right)\\
&C=\bigoplus_{i\geq 0}\left([i+1,0,0,0,0]+[i,0,0,0,2]++[i,0,1,0,0]\right)
\end{split}
\end{equation}
Then
\begin{equation}
\begin{split}
&H_{0}(V,M_{0}\otimes M_{0})=A+\mathbb{C}+\Lambda^2(V)+\Lambda^4(V)\\
& H_{0}(V,M_{0}\otimes M_{0})=A+V+\Lambda^3(V)+\Lambda^5(V)\\
&H_{i}(V,M_{0}\otimes M_{0})=\Lambda^{i+4}(V)\quad i\geq 2\\
&H_{0}(V,M_{0}\otimes M_{1})=B+[0,0,0,0,1] \quad H_{1}(V,M_{0}\otimes M_{1})=B+[0,0,0,1,0]\\
&H_{i}(V,M_{0}\otimes M_{1})=0\quad i\geq 2\\
&H_{0}(V,M_{1}\otimes M_{1})=C \quad H_{1}(V,M_{1}\otimes M_{1})=C+\mathbb{C}\\
& H_{i}(V,M_{1}\otimes M_{1})=0\quad i\geq 2
\end{split}
\end{equation}
\end{proposition}

\begin{proposition}\label{P:ududufgqex}

$H^0(YM,IU(TYM))=0, H^1(YM,IU(TYM))=[1,0,0,0,0]+[0,0,0,1,0]$. The elements of $[1,0,0,0,0]$ have degree two, of $[0,0,0,1,0]$ degree one
\end{proposition}

\begin{proof}
The universal enveloping algebra $U(TYM)=T(M)$ admits a filtration by powers of the augmentation ideal $I\subset U(TYM)$. The adjoint action of $YM$ preserves $I$, hence the filtration $F^i=I^{\times i}$. We plan to compute cohomology $H^i(YM,U(TYM))$ using a spectral sequence of mentioned filtration. 

The $E_2$ term of it  is equal to $E^{ij}_2=H^{i+j}(YM,M^{\otimes j})$. We computed $H^1(YM,M^{\otimes j})$ for $j\geq 1$. According to proposition (\ref{P:qsfwqcc})  the groups are equal to zero for $j\geq2$. We examine  the differential in the spectral sequence on the group $H^1(YM,M)$. The differential $\delta$ acts:
\begin{equation}\label{E:kdjvsh}
\delta:H_2(YM,M)\rightarrow H_1(YM,\Lambda^2[M])\subset H_1(YM,M^{\otimes 2})
\end{equation}
$H^0(YM,IU(TYM))=0$ due to proposition (\ref{P:udhgewgfa}).

In the following part of the section we formulate essential lemmas needed for the  proof that the kernel of $\delta$ is equal to $V+S^*$.
\end{proof}

\begin{proposition}
The map $B_1:H_1(V,M\otimes M)\rightarrow H_2(YM,M)$ is surjective.
\end{proposition}
\begin{proposition}\label{P:mxbste}
The map $\delta$ is an embedding on the image $\Im(H_1(V,\Lambda^2(M))\subset H_2(YM,M)$. The composition $\delta^c\circ \delta:H_2(YM,M)\rightarrow H_2(YM,M)$ is a projection on $\Im(H_1(V,\Lambda^2(M))$
\end{proposition}
\begin{definition}\label{D:whysbxg}
For a (graded) Lie algebra $\mathfrak{g}$ the group $D(\mathfrak{g})$ is a group of inner $\mathfrak{g}$-invariant co-products:
 $D(\mathfrak{g})=\Sym^2(\mathfrak{g})_{\mathfrak{g}}$. The linear space $D(\mathfrak{g})$ is generated by elements $a\circ b, a,b\in\mathfrak{g}$. Subject to relation $[a,b]\circ c+b\circ [a,c]=0, a\circ b=b\circ a$ and the symbol is linear with respect to each of the arguments.
\end{definition}
We would like to specialize  construction of definition (\ref{D:whysbxg}) to  algebra \\ $\mathfrak{h}=TYM/[TYM,[TYM,TYM]]$. The algebra $\mathfrak{h}$  is a direct sum of two linear spaces $M+\Lambda^2(M)$. The linear space $\Lambda^2(M)$ is  the center . The commutator $[.,.]:M\wedge M\rightarrow \Lambda^2(M)$ is an isomorphism.

\begin{proposition}
A linear space $D(\mathfrak{h})$ is a $\Sym(V)$-module. There is a short exact sequence of modules
\begin{equation}
0\rightarrow \Lambda^3(M) \rightarrow D(\mathfrak{h})\rightarrow \Sym^2(M)\rightarrow0
\end{equation}
\end{proposition}
\begin{proposition}\label{P:lsjshcgs}
There is a commutative diagram
\begin{equation}\label{E:jssux}
\mbox{
$\begin{array}{ccc}
H_2(YM,M) & \overset{\delta}{\rightarrow} & H_1(YM,M^{\otimes2})\\
\uparrow B_2& &\uparrow B_3\\
H_1(V,\Sym^2(M)) & \overset{\tilde {\delta}}{\rightarrow} & H_0(V,\Lambda^3(M))
\end{array}$
}
\end{equation}
The map $\tilde \delta$ is the boundary map corresponding to extension $D(\mathfrak{h})$.
The map $B_3$ in the diagram above has a trivial kernel. The kernel of map $B_2$ in  (\ref{E:jssux}) has kernel equal to $\Lambda^5(V)$.
\end{proposition}
\begin{proposition}
The kernel of  $\tilde {\delta}$ is $\Lambda^5(V)+V+S^*$.
\end{proposition}
\begin{proof}
We send the reader to \cite{M3}. An important comment is that a nontrivial $Spin(10)$-invariant cocycle $a\in H_1(V,M_1\otimes M_1)$ is graded anti-symmetric and can be ignored in the present discussion.
\end{proof}

\begin{proposition}
Elements of $[0,0,0,1,0]\subset H^{1,1}(YM,TYM)$ can not be lifted to $ H^{1,1}(L,TYM)$
\end{proposition}
\begin{proof}
According to long exact sequence (\ref{E:ywers}) the linear space $ H^{1,1}(L,TYM)$ is equal to zero.
\end{proof}

\end{document}